\documentclass[12pt,preprint]{aastex}

\usepackage{lscape}

\shorttitle{Frequency Standards for VLBI at the Highest Frequencies}
\shortauthors{Rioja et\,al.}

\begin{document}

\title{The Impact of Frequency Standards on Coherence in VLBI at the Highest Frequencies} 

\author{}
\author{M. Rioja\altaffilmark{1,2}, R. Dodson\altaffilmark{1}, Y. Asaki\altaffilmark{3,4}, 
J. Hartnett\altaffilmark{5}, S. Tingay\altaffilmark{6}}
\affil{$^1$ICRAR, University of Western Australia, Perth, Australia; 
$^2$Observatorio Astron\'omico Nacional (OAN), Apartado 112, E-28803,  Alcala de Henares, Espa\~na; \\
$^3$Institute of Space and Astronautical Science, 3-1-1 Yoshinodai, Chuou, Sagamihara, Kanagawa 252-5210, Japan \\
$^4$Department of Space and Astronautical Science, School of Physical Sciences, The Graduate University of Advanced Studies (SOKENDAI), 3-1-1 Yoshinodai, Chuou, Sagamihara, Kanagawa 252-5210, Japan\\
$^5$School of Physics, University of Western Australia, Perth, Australia\\
$^6$ICRAR, Curtin University, Perth, Australia }

\email{maria.rioja@icrar.org}


\begin{abstract}

  We have carried out full imaging simulation studies to explore the
  impact of frequency standards in millimeter and sub-millimeter Very
  Long Baseline Interferometry (VLBI), focusing on the coherence time
  and sensitivity. In particular, we compare the performance of the
  H-maser, traditionally used in VLBI, to that of ultra-stable
  cryocooled sapphire oscillators over a range of observing
  frequencies, weather conditions and analysis strategies.  Our
  simulations show that at the highest frequencies, the losses induced
  by H-maser instabilities are comparable to those from high quality
  tropospheric conditions.  
We find significant benefits in replacing H-masers with cryocooled
sapphire oscillator based frequency references in VLBI observations at
frequencies above 175 GHz in sites which have the best weather
conditions; at 350 GHz we estimate a 20-40\% increase in sensitivity,
over that obtained when the sites have H-masers, for coherence losses of 20-10\%, respectively.
  Maximum benefits are to be expected
 by using colocated Water Vapour Radiometers for atmospheric
  correction. In this case, we estimate a 60-120\% increase in
  sensitivity over the H-maser at 350 GHz.

\end{abstract}

\keywords{Instrumentation: interferometers -- Instrumentation: miscellaneous}

\section{Introduction}

Pushing Very Long Baseline Interferometry (VLBI) observations to
higher frequencies is a desirable approach to increase angular
resolution and to target new and unique fields of science
investigation.  Nevertheless, the observations become increasingly
challenging as the wavelengths get shorter.  VLBI at millimeter and
sub-millimeter wavelengths ({\it hereafter} mm-VLBI) suffer from low sensitivity which makes
operations extremely difficult and dramatically restrict the scope of
application. Thus mm-VLBI observations have been very limited.  While
wider bandwidths and increased recording rates will help alleviate the
problems, the major breakthrough to improve sensitivity will come from
directly addressing the causes.  The sensitivity issues have their
origin in the large coherence losses induced by the fast phase
fluctuations imposed on the astronomical signal, which arise from
atmospheric propagation of the radio waves and instrumental noise
contributions, such as from the frequency standards \citep{tms}.
These fast fluctuations impose short coherence times, which in turn
increases the minimum detectable flux.  To a lesser extent, the
attenuation of the signal from atmospheric transmission also
contributes.  Moreover, in the high frequency regime the sensitivity
of radio telescopes is reduced, the receiver noise is increased, and
the intrinsic source strength is generally lower.

The reason for the atmospheric propagation issues lies in the
variations in its refractive index, mainly due to 
inhomogeneities in the distribution of tropospheric water vapour in 
the turbulent atmosphere.
These variations result in changes in the electrical path
length through the atmosphere, which introduces phase fluctuation errors
on spatio-temporal scales in the observed phase of the astronomical
signal, degrading the sensitivity by reducing the coherence time.
The fractional frequency stability contribution of a typical atmosphere to the
received signal has been determined to have an Allan 
Standard Deviation (ADEV) of $\sim$10$^{-13}$s/s 
up to 100 seconds \citep{rogers, rogers2}.
Given the non dispersive nature of the troposphere, the induced
temporal phase fluctuations scale linearly with observing
frequency.
Hence, to mitigate the fundamental limitation imposed by 
atmospheric propagation, it is essential to select a site with
exceptionally stable and transparent
atmospheric conditions;
dry high altitude sites.
In addition, the potential of using inferred corrections 
estimated from measurements of sky brightness temperatures using co-located
Water Vapour Radiometers (WVR) is a long term on-going effort.
Currently, the ALMA (Atacama Large Millimeter Array \footnote{http://www.almaobservatory.org}) project is developing state-of-the-art
WVRs to compensate for tropospheric fluctuations.
Preliminary observations with these instruments show promising performance, with progress towards
their nominal specification (98\% compensation of the fluctuations) \citep{nic_wvn}. 
WVR's application to VLBI holds the promise
of a significant breakthrough in dealing with current tropospheric
limitations \citep{honma_alma,matsushita_spie}.

In mm-VLBI, the other dominant error contribution to the residual phase
fluctuations arises from the instabilities 
of the frequency standard traditionally used in VLBI, the Hydrogen
(H)-maser.  
The aim of the work presented here is to assess the benefits of
replacing a H-maser with an ultra-stable cryoCooled Sapphire
Oscillator (CSO), whose stability is more than 1 order of magnitude
better at the short timescales of interest in this field, as an option
for mm-VLBI.

Previous studies of Doeleman and collaborators investigated a
liquid helium cooled Sapphire Oscillator, along with the estimated performance of 
a 10 MHz synthesized signal, and showed a moderate benefit at 350 GHz. 
The next generation of CSOs has become a very robust, low
maintenance, long term operational alternative to the H-maser as an
ultra-stable frequency standard, now that the device operates with a
low vibration pulse-tube cryocooler \citep{wang10,Hartnett10,HartnettandNand10}.
The CSO's performance is well established and the measured frequency
stability for both short and long term timescales is
better than that achieved with the liquid
helium cooled version \citep{Doeleman11,nand11,hartnett12}.
Additionally, the new version generates a 100-MHz synthesized signal
which has significantly better performance \citep{nand11}. 
The Doeleman study was completed before these improvements to the CSO
were implemented and were largely analytic, being based on the standard
expressions for coherence. We have completed full imaging
simulations which are much more sensitive to the cumulative
effects of different error sources. Furthermore, only a limited analysis of the effect of 
WVR was included in Doeleman's studies, which we have considerably advanced.

In this paper we present the results from full imaging simulations to 
explore the performance of VLBI observations using a H-maser versus a
CSO under a variety of conditions, to answer
the questions of whether, and under
which circumstances, the CSO could be expected to provide benefits over a H-maser. 
Our simulation studies also explore the 
potential of various analysis strategies to enhance the coherence time, in
particular using simultaneous dual frequency observations 
and the use of collocated WVRs to correct for tropospheric fluctuations.

This is the first step in the process to fully test the application
of ultra-stable CSOs to facilitate mm-VLBI observations.
The results obtained from these simulation studies will serve as a 
guideline for an empirical demonstration with actual observations.
In this paper, a description of the simulation procedure is presented in Section
2. The results are presented in Section 3, and discussion of results in Section
4.

\section{The Method}\label{sec:method}

In this section we describe the procedure followed in our full imaging
simulation studies to explore the
benefits of replacing a H-maser with a CSO,
along with the potential of different analysis strategies, in mm-VLBI.
First, we used an updated version of ARIS \citep{A07} to generate the
synthetic datasets 
which replicate the 
interferometric visibilities obtained from a correlator
for a given
observing configuration, i.e. coordinates of the telescopes
and the celestial target, observing frequencies, bandwidth,
atmospheric conditions, etc.
ARIS is a versatile software tool which includes complete two
dimensional geometric
and atmospheric phase screen models, and has been used in previous simulation
studies for ground and space VLBI \citep{A07, rioja_vsop2}.  We have
found it reproduces the results from actual observations extremely
well \citep{asaki_10}.  Among the functionalities added in order to carry out this
project are the options to incorporate the noise contributions from
the station frequency standards (i.e. clock errors), for both H-masers
and CSOs, to select ALMA-site type weather conditions (i.e. Very Good),
and to generate WVR corrected tropospheres, in simulations at
frequencies up to 350 GHz.

Table \ref{tab:one} lists the parameter space explored in our simulations.
These comprise of observations at frequencies between 86 to
350 GHz, using H-maser and CSO clocks, for a range of weather
conditions classified as: Very Good, Good, Typical and Poor, 
using different telescope networks, and realistic errors 
for the model parameters.
We also explore the case of having WVR-corrected atmospheres.
These simulations are mainly concerned with the relative impact
of tropospheric and clock instabilities. Therefore, in order to avoid bias 
effects introduced by observations of weak sources, the simulations 
were all run with strong signals and are thermal noise free.

\begin{figure}
\includegraphics[width=12cm]{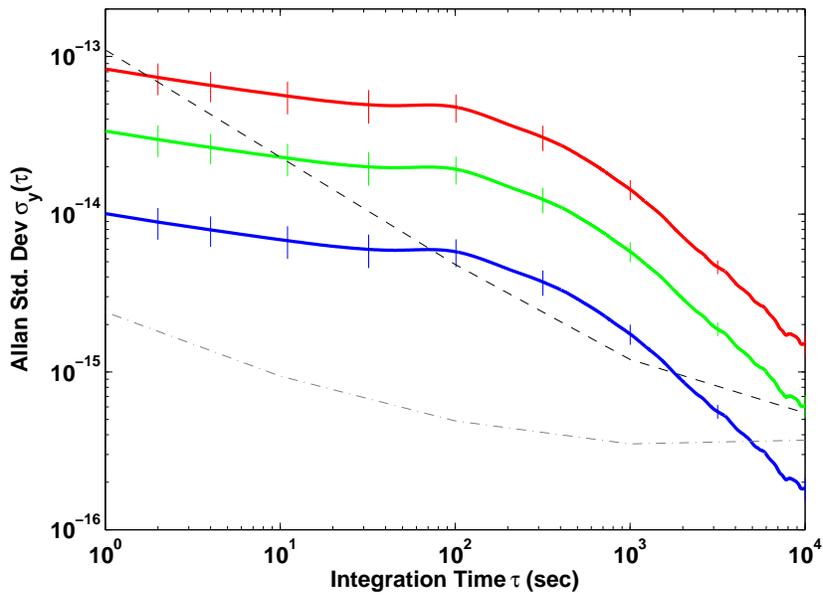}
\caption{Plot of the Allan Standard Deviations for the atmospheres and
  the frequency standards from ARIS. Shown in solid lines are the Good (in red,
  upper most), the Very Good (in green) and the WVR-corrected (blue, lowest) atmospheres. The dash line is the ADEV for a KVARZ H-maser and the dot-dash for the CSO-100 MHz. The atmosphere values are extracted from the mean of repeated runs of independent model generation in ARIS, with error bars marking the $\pm$1-$\sigma$ standard deviation from the mean.
}\label{asd}
\end{figure}

ARIS implements the clock and tropospheric fluctuations by generating a time series 
of random values which are statistical realizations of
models in the code. 
The functional forms of the time standard noise used are 
polynomial expansions of the ADEV
which follow Nand et~al. (2011)  
and represent the observed CSO behaviour and the
quoted  behaviour of the Russian KVARZ H-maser. 
The simulated fluctuations of the atmospheric path can be generated
for a choice of four weather conditions, i.e. Very Good (V), Good (G), Typical (T)
and Poor (P), adjusting the modified coefficient of the spatial
structure function of the troposphere to values 0.5, 1, 2 and $4
\times 10^{-7} m^{-1/3}$, respectively, with the assumption of
Kolmogorov turbulence \citep{A07,beasley}.  For example, the ``Very
Good'' weather corresponds to an {\sc rms} excess path length of 100
microns for a 100-m baseline (ca. 50-percentile tropospheric condition at Chajnantor Observatory site (Chile)) 
\citep{asaki_m535}.
Larger fluctuations correspond to more
unstable tropospheric conditions.
The option for WVR phase correction (W) implemented in ARIS matches
empirical results from ALMA commissioning and science verification
observations, with 
ALMA WVRs at 183 GHz (for 600-m baselines) \citep{matsushita_spie},
which typically result in the {\sc rms} phase reduced to 
30-40\%. Hence, the W option reduces the {\sc rms} phase of 
Very Good weather by a factor of 3, keeping the assumption of Kolmogorov turbulence
and adds a WVR path length measurement Gaussian noise to each 
antenna (1-$\sigma$ equal to 4 microns, for 10 seconds average data). 

Figure \ref{asd} shows the model frequency stability,
characterized in terms of the ADEV, 
for the clock (for CSO and H-maser) and atmospheric (W, V, and G) errors 
with ARIS.
We have confirmed that the behaviour of the atmospheric models in
ARIS are consistent with known approximations and published
results. The driver for our detailed examination was that the
coherence times we derived appeared to be in excess of the
established limits.  These resolved themselves into differences in the
definition of loss limits deemed acceptable or in the procedures
followed.
Different authors use different cutoffs for the coherence losses, 
and these lead naturally to different coherence times. We have used 
the maximal value for the allowed phase {\sc rms} of 1 radian, corresponding to an 
approximately 40\% 
signal loss, following the discussions in \citet{tms} (Eq. 9.119). 
On the other hand, setting limits of 10\% signal loss translate into
a phase {\sc rms} of 0.4 radians.
We have calculated the coherence factor at 86\,GHz with Eq. 7 of
\citet{nand11} assuming the standard ADEV of `Good' tropospheric
conditions. We find identical losses as a function of integration time 
as those derived from the ARIS model atmospheres.
The matching coherence times were obtained both by direct averaging, and the similar approach 
of phase-only calibration (with CALIB in AIPS \citep{aips}).

We have performed all our simulations in the strong signal domain,
assuming that the next generation telescopes making up the Event
Horizon Telescope \citep{eht}
will be highly sensitive with excellent performance and wide
bandwidths. Therefore we used a 10\,Jy model source and no thermal
noise was added. In the strong signal domain one can use coherent
methods (i.e. fringe finding for delays and rates as well as phase)
and direct imaging, as opposed to forming closure products with
incoherently derived delays and model fitting the data. This allows us
to extend the coherence time well beyond the values obtained following
the methods above. As the ADEV of our model atmospheres follow those
of observed atmospheres, we are confident that fringe-fitted
real data will show similar behaviour. We find our simulated results
compare with the coherence quoted in papers on VLBI observations at 86
GHz (e.g.  \citet{enno_fpt}).

The simulated datasets 
comprise all combinations of tropospheric conditions and clock types for observations at each
frequency (i.e. 86, 175 and 350 GHz), and for simultaneous dual-frequency 
observations (i.e. 43/86GHz; 87/175GHz; 175/350GHz).
In addition, datasets with a single noise contribution,
either atmospheric or clock errors, have been used 
to explore the individual signatures and their dependence on observing frequency. 
Table \ref{tab:two} summarizes the 5 simulation cases presented in
this paper.
The output of ARIS are standard {\it uv}-FITS files with 
corrupted visibilities, as described in Table \ref{tab:two} for each configuration, 
ready for processing with AIPS.

Each synthetic dataset was analysed using calibration and imaging procedures 
in AIPS to produce images.
The analysis was carried out
following different calibration strategies, listed in Table \ref{tab:two}: \\
1) standard VLBI self-calibration procedures for the analysis of single
frequency datasets (i.e. Cases 1,2,3). For each dataset, we repeated the analysis
with different FRING solution intervals in the range from 0.1 to 6 minutes 
to explore coherence losses;\\
2) frequency phase transfer calibration (FPT) procedures \citep{riojadodson11} for
the simultaneous dual-frequency datasets. That is, the high frequency data are
calibrated using the scaled self-calibration estimates from the lower
frequency  (i.e. Case 4). The FPT calibration
compensates all non-dispersive errors, but long timescale ionospheric
errors, and in general any dispersive error, still remains. \\
3) a hybrid calibration with FPT and self-calibration 
(i.e. Case 5). That is, same as 2) combined with a further cycle of self-calibration
with a much longer solution interval in FRING to correct for the dispersive terms.

In all cases, the calibrated complex visibility samples were Fourier
transformed, 
and deconvolved, using standard imaging procedures with the AIPS task IMAGR, to produce an image.
And finally, the map peak flux 
was measured using the AIPS task JMFIT. 
The fractional flux recovery (FFR) quantity, defined as the ratio
between the peak flux in the map divided by the source model
flux (in ARIS), was used as the figure of merit to 
evaluate and compare the coherence losses 
for each configuration.
The values presented here are the mean 
and {\sc rms}
of the FFR measured
from 5 simulation runs for each configuration, each with 
independently generated noise time  series 
for the model errors in ARIS. 
Larger values of the FFR correspond to smaller
coherence losses, which in turn lead to higher sensitivity.  This comparison
is ultimately used to assess the benefits of replacing the H-maser with
CSOs.

\begin{table}
\caption{Parameter space of our simulations with ARIS} \label{tab:one}
\begin{tabular}{ll}
\hline
{\bf GEOMETRY$^{(1)}$:} & \\
\hspace*{0.1cm} {\it Array}  & VLBA\\
\hspace*{0.1cm} {\it Source} & Compact and Strong \\ \hline
{\bf PROPAGATION MEDIA$^{(2)}$:} &  \\
\hspace*{0.1cm} {\it Ionosphere} & Error: 6 TECU \\
           & Fluctuations: Nominal \\
\hspace*{0.1cm} {\it Troposphere} & Error: 3 cm \\
            & Fluctuations: WVR-corrected, V, G, T, P \\ \hline
{\bf FREQUENCY STANDARD:} & H-maser, CSO-100 MHz \\ \hline
{\bf OBSERVING FREQUENCY:} &  \\
\hspace*{0.2cm} {\it Single Frequency} &  86, 175, 350 GHz \\
\hspace*{0.2cm} {\it Dual Frequency} &  43/86 GHz, 87/175 GHz, 175/350 GHz \\ \hline
\end{tabular}

\vspace*{1cm}

{\footnotesize {\bf (1):} {
VLBA: Very Long Baseline Array. 
The telescope and source coordinates are simply used to generate the
tracks, i.e. sampling of the uv-plane. } \\
{\bf (2):} See Asaki et al. (2007) for detailed information. }
\end{table}

\begin{table}
\caption{Description of the simulation study cases presented in this paper, based on 
 the nature of the error contributions to the synthetic datasets generated with ARIS, and
the data analysis procedure with AIPS.}\label{tab:two}

\begin{tabular}{|ll|c|c|c||c|c|}
\hline
  & & \multicolumn{5}{c|}{\bf Simulation Cases}  \\
  \cline{3-7}  &  & Case 1 & Case 2 & Case 3 & Case 4 & Case 5 \\ \hline
 \multicolumn{7}{l} {{\bf ERRORS}}  \\ \hline 
 & {\rm \underline{GEO errors}} & & & X & & \\ 
  & {\rm \underline{CLOCK noise}} &  & &  & & \\
    & \hspace{0.2cm}{\it H-maser; CSO} & X & & X &  X &  X \\ 
 & {\bf \underline{ATM noise}} & &  &  & & \\
 & \hspace{0.2cm}{\it V;G;T;P} & & X & X & X & X \\ 
& \hspace{0.2cm}{\it WVR-corrected (W)} & & X  & X & & \\
 & {\rm \underline{Obs. FRQ (GHz)}} & 86 & 86 & 86 &  43/86 & 43/86 \\
 &     & 175 & 175 & 175 & 87/175 & 87/175 \\
 &     & 350 & 350 & 350 & 175/350 & 175/350 \\
\hline \hline
 \multicolumn{7}{l} {\bf ANALYSIS}  \\ \hline
 & \underline{SC}$^1$ & X & X & X & & \\
 & \underline{FPT}$^2$ & & &  & X & \\
 & \underline{FPT + SC}$^3$ & & & & & X \\
\hline
\hline
\hline 
\end{tabular}

\vspace*{1cm}

{\footnotesize {\bf (1):} {Self-Calibration}\\
{\bf (2):} {Frequency Phase Transfer, using dual-frequency observations.}\\
{\bf (3):} {A combination of the two above. }}
\end{table}

\section{Results}

The final products of our simulations are images, and the FFR values
measured from them.  We use FFR as the figure of merit to compare
performance, that is the associated coherence losses, for the
simulated configurations.  In this section we present the results from
our simulations following a classification similar to that
shown in Table  \ref{tab:two}. 
The interpretation of these results in terms of sensitivity is addressed in
section 4.

\subsection{{\it Case 1: Clock Errors Only (Atmosphere Noise-Free)}}

\begin{figure}
\includegraphics[width=8cm]{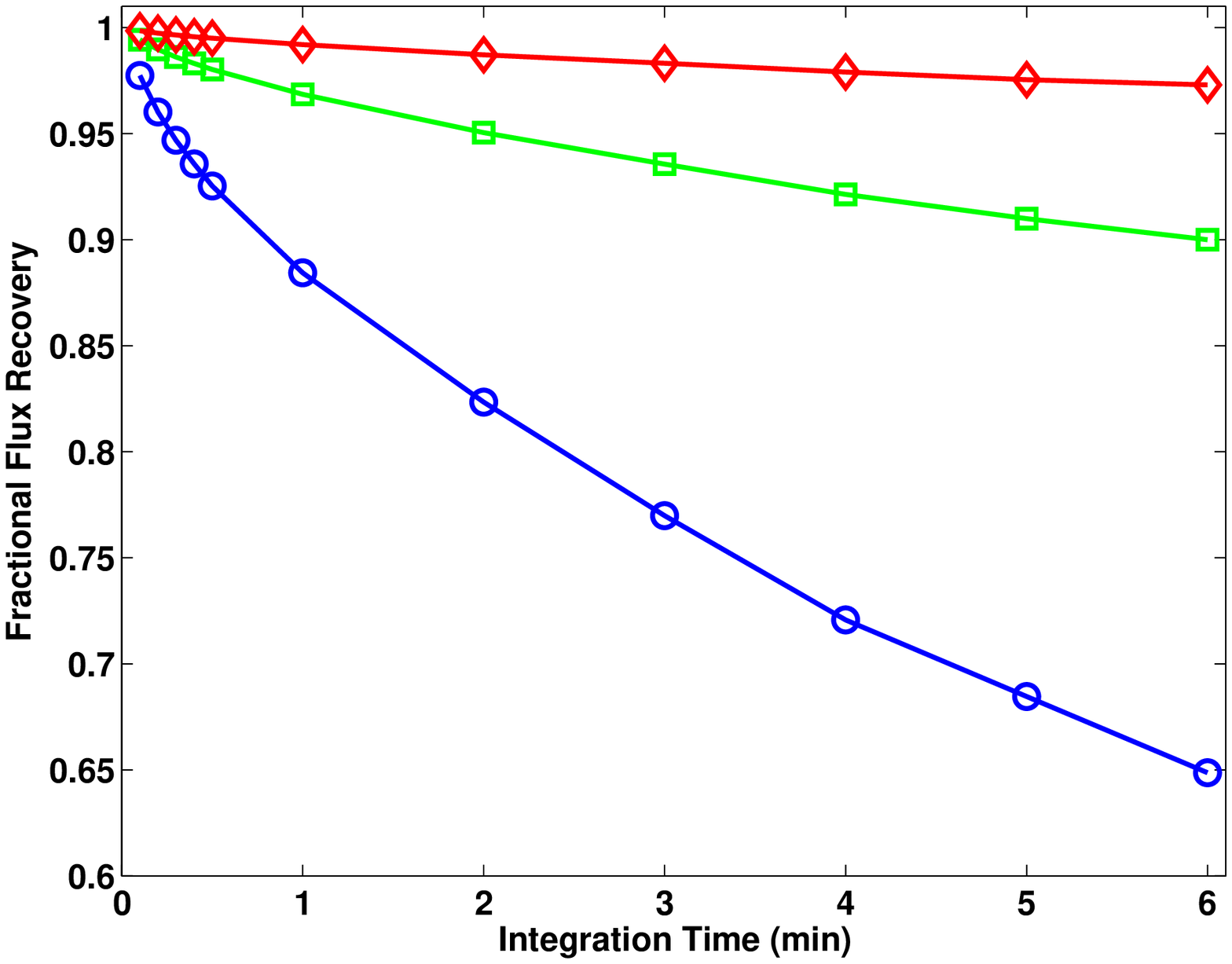}
\includegraphics[width=8cm]{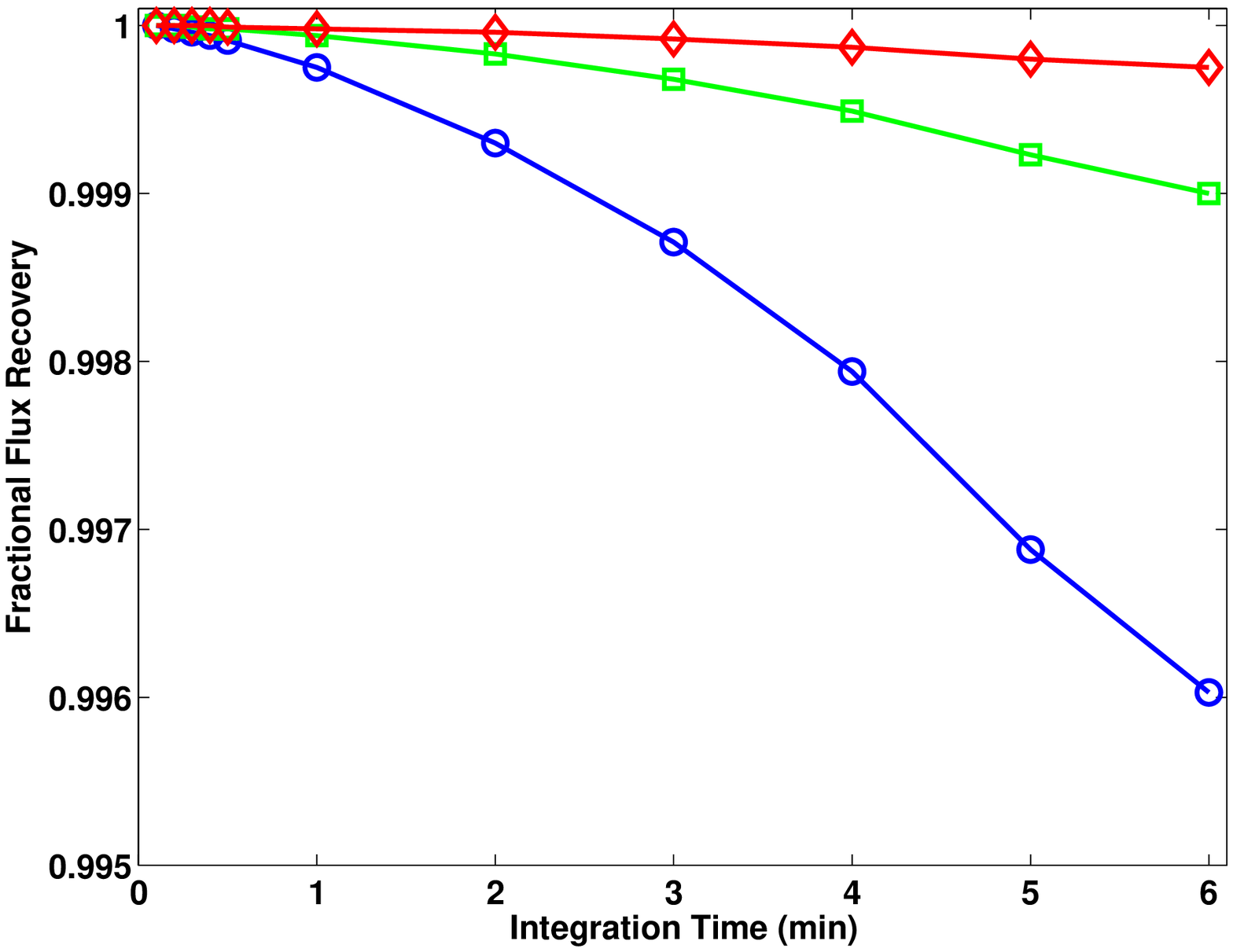}
\caption{{\it Left:} Coherence losses, plotted as the Fractional Flux Recovered, introduced by the H-maser clock instabilities 
as a function of the integration time, in observations at 86 (red
diamond), 175 (green square) and 350 
(blue circle) GHz; {\it Right:} same, for CSO-100 MHz.}
\label{case1}
\end{figure}

Figure \ref{case1} shows the FFR values from simulations with clock
noise only at 86, 175 and 350 GHz, for a range of integration times in
the self-calibration analysis with AIPS.
The coherence losses associated
with the H-maser instabilities significantly increase with the observing
frequency, and for a given frequency at longer integration times.  In
comparison, the coherence losses from the CSO are negligible, at all
frequencies and for the same range of integration times.
Our simulations with clock noise only show that the instabilities from a
H-maser result in $\sim 10\%$ losses in 0.5-1 minutes at 350 GHz, and
$\sim 20\%$ losses in $\sim 2$ minutes. The corresponding losses for the 
same integration times in the case of the CSO are $\le 0.5\%$.\\

\subsection{{\it Case 2: Atmospheric Errors Only  (Clock Noise-Free)}}

\begin{figure}
\includegraphics[width=8cm]{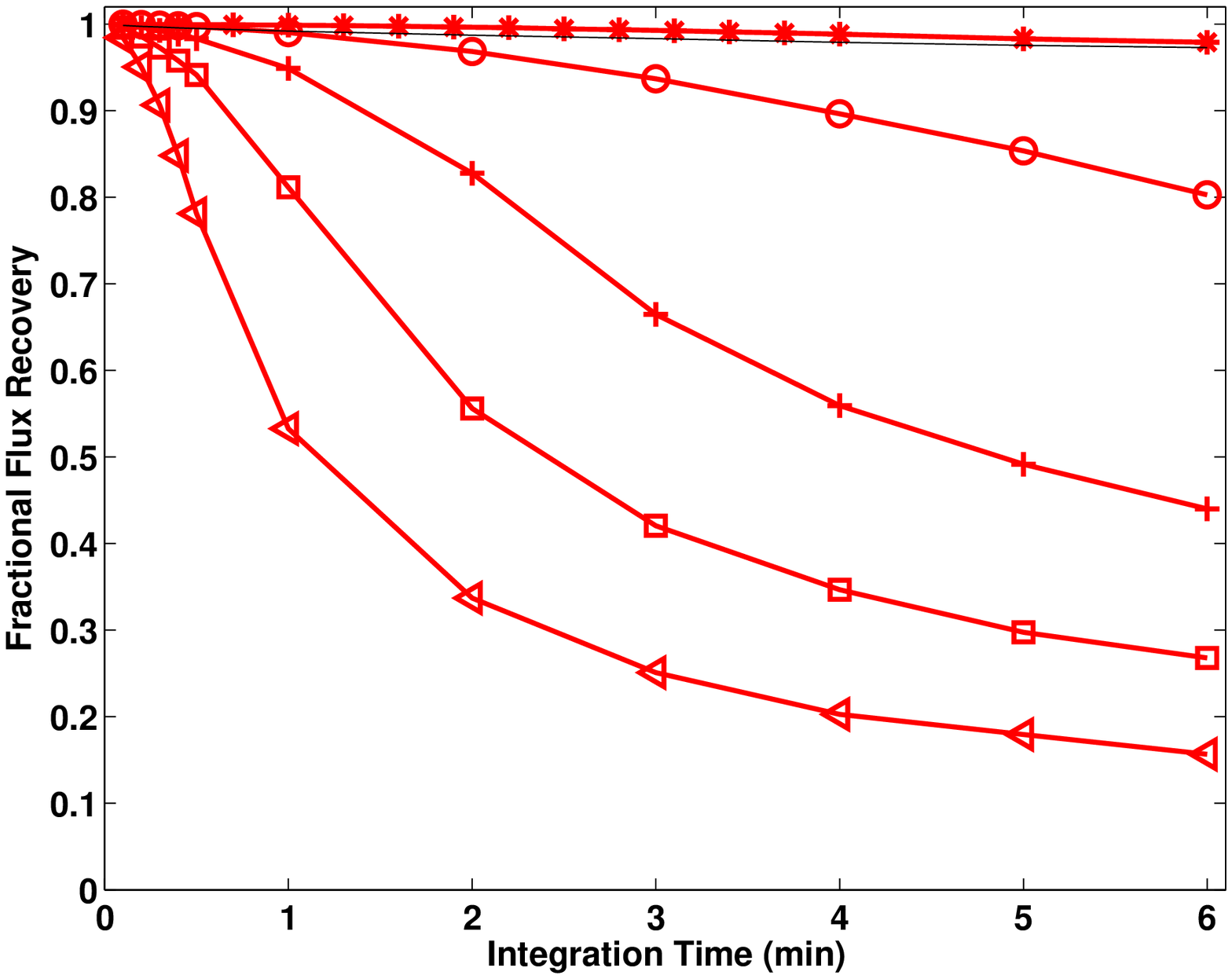}
\includegraphics[width=8cm]{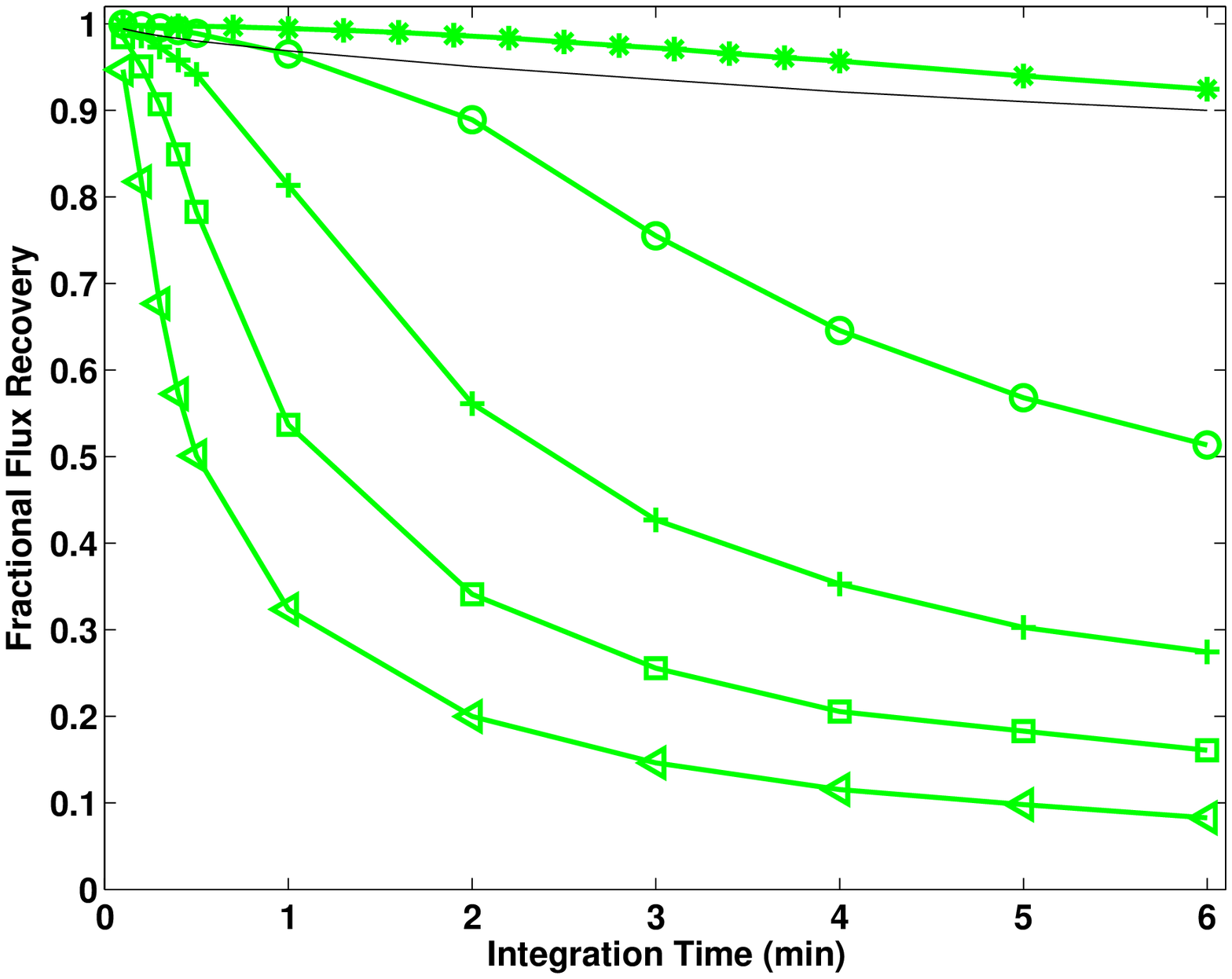}
\includegraphics[width=8cm]{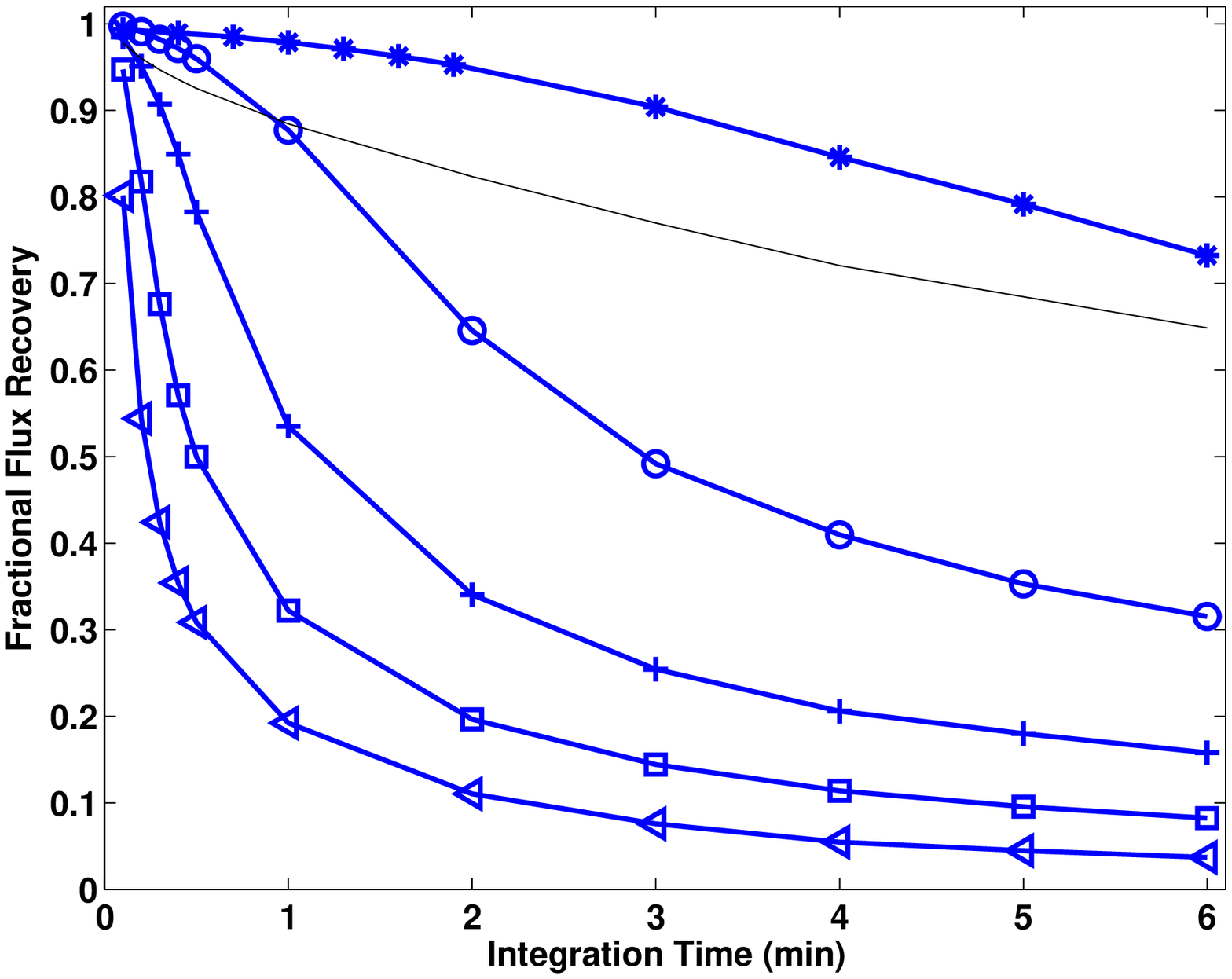}
\caption{Atmospheric-induced coherence losses, plotted as the Fractional Flux Recovered, versus integration time estimated for a range of 
 weather conditions (V (circle), G (plus), T (square), P (diamond))
 and for WVR-corrected atmospheres, W (star), at 86 ({\it upper left}), 
175 ({\it upper right}) and 350 ({\it lower left}) GHz. At each frequency, the sequence of lines starting from 
the uppermost correspond to W,V,G,T,P weather conditions. The {\sc rms} errors are smaller than the symbol size.
For illustration, a light black line indicates the losses due entirely to a H-maser (shown in Figure 2).}
\label{case2}
\end{figure}

Figure \ref{case2} shows the results from simulations involving only atmospheric
instabilities (i.e. the clock-noise free case).  
At a given frequency, the FFR decreases with increasing integration times,
and with worsening weather conditions, being highest for V and lowest for 
P for each integration time value, as expected from increasing signal coherence losses. 
For example, at 350 GHz these simulations result in peak flux losses of $\sim 20\%$ on
timescales of $\sim 1$ minute, under V weather conditions, and
$\sim 0.5$, 0.2 and 0.1 minutes with G, T and P weather
conditions, respectively. 
Also, for a given weather and integration time,  
the coherence losses are increasingly large at higher frequencies.
This is in agreement with the well 
recognized importance of selecting sites with excellent weather 
conditions for VLBI at the highest frequencies.  
Best results at all frequencies are found when using WVR-corrected atmospheres; 
for example, at 350 GHz, these simulations result in peak flux losses of $\sim 20\%$ on much longer
timescales of $\sim 5$ minutes. \\

\subsection{{\it Comparison of Case 1 and Case 2}} 

\begin{figure}
\includegraphics[width=8cm]{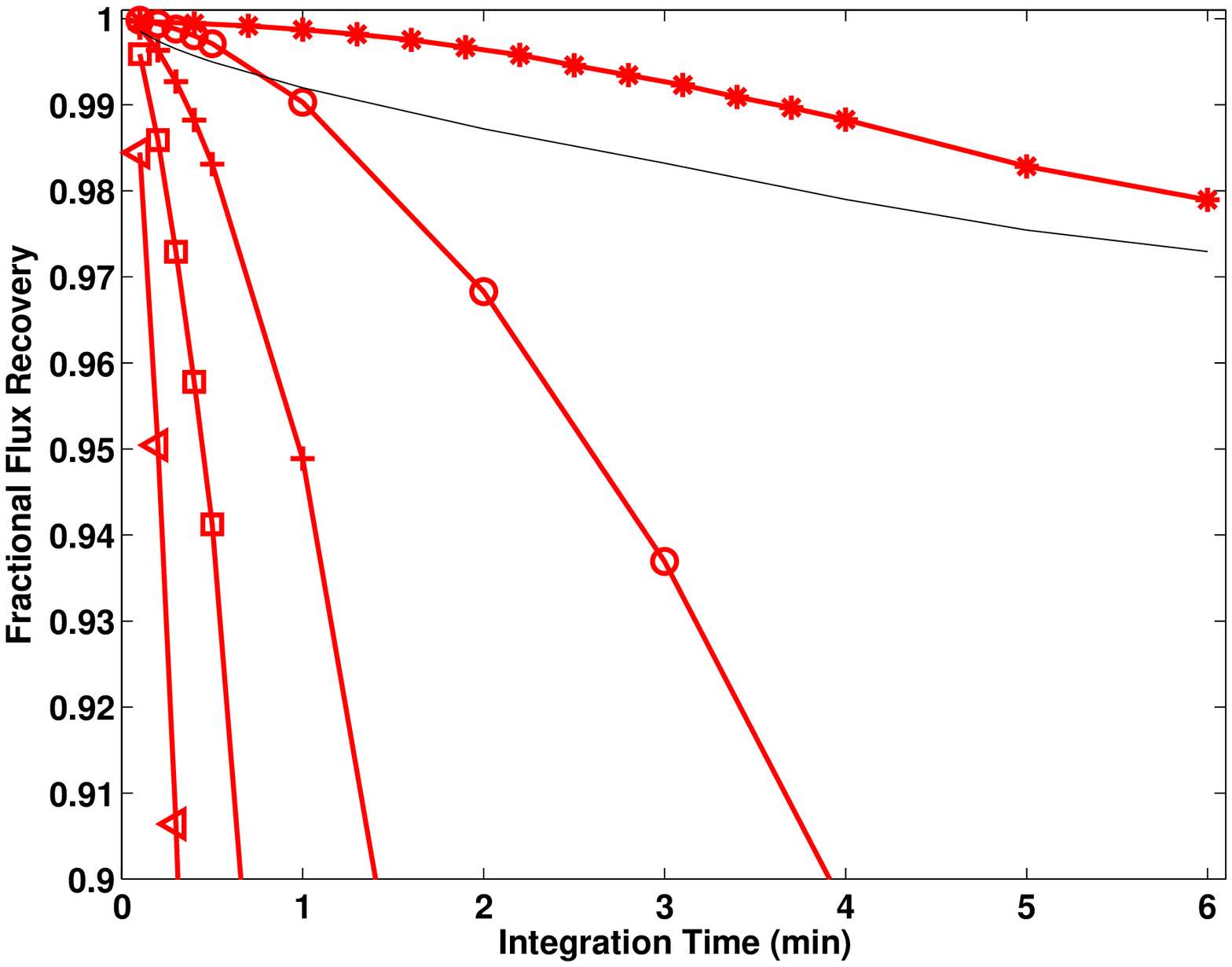}
\includegraphics[width=8cm]{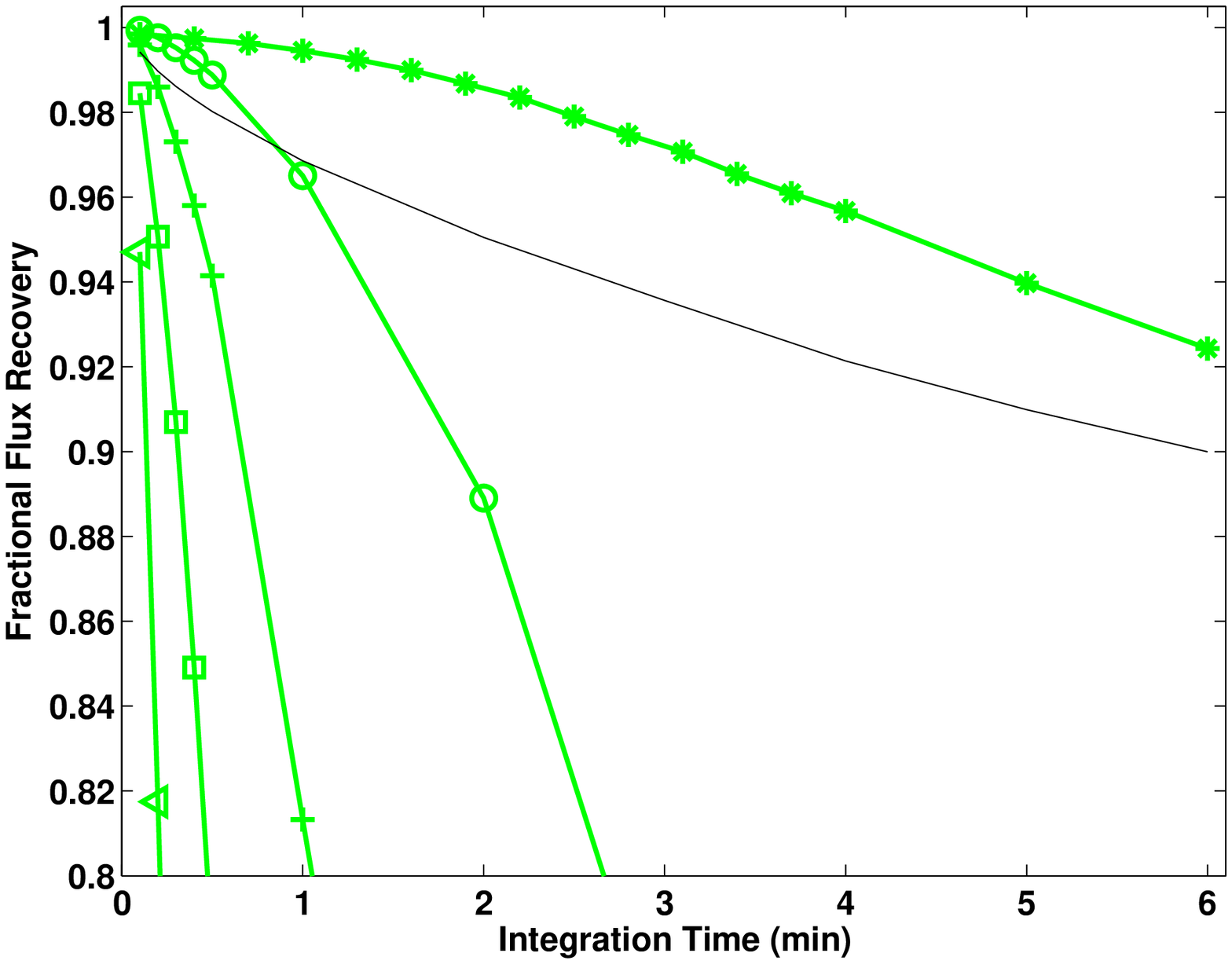}
\includegraphics[width=8cm]{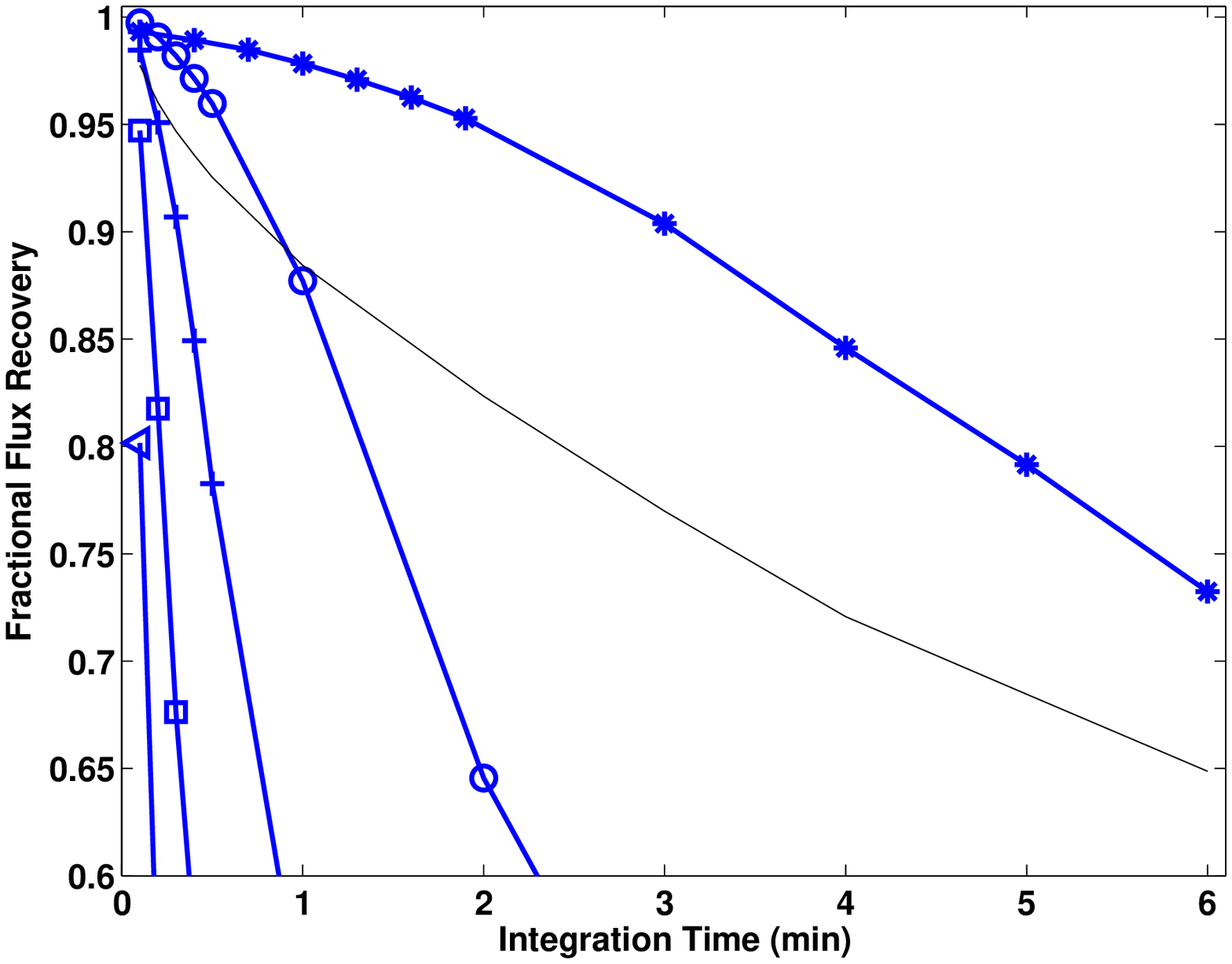}
\caption{Zoomed versions of Figure 3, focusing in on the short
  timescales at which the H-maser losses are greater than, or a
  significant fraction of, the atmospheric contributions. As in Figure 3,
  the atmospheric-induced coherence losses (the Fractional Flux Recovered) 
  estimated for a range of weather conditions (V,G,T,P) and for
  WVR-corrected atmospheres (W), are plotted versus integration time
   at 86 ({\it upper left}), 175 ({\it upper right}) and 350 ({\it
     lower left}) GHz. At each frequency, the
  sequence of lines starting from the uppermost correspond to
  W,V,G,T,P weather conditions. A light black line indicates the
  losses due entirely to a H-maser.}
\label{case1-2}
\end{figure}

Figure \ref{case1-2} shows the superimposed results 
from the simulation cases above: with only clocks and with only
tropospheric noise contributions.
The point at which the losses from the H-maser and the tropospheric
instabilities are equal increases in significance with the observing
frequency and the quality of weather conditions.  
For sufficiently short timescales ($<$1
minute) the H-maser losses dominate under V weather conditions, with
contributions of
up to 15\% at 350 GHz, but are negligible at 86 and 175 GHz;
these timescales are much longer ($>$6 minutes)
for WVR-corrected atmospheres, with H-maser losses up to $>$10\% and 35\%, at 175 and 350 GHz, 
respectively. 
Note that the coherence losses with the CSO-only are negligible ($ < 0.5 \%$) in all
cases, and have not been included in the plots. 

\subsection{{\it Case 3: Clock and Atmospheric Noise}}

In complete simulations, with joint tropospheric and clock
instabilities, the estimated losses are the combination of the
corresponding individual ones. Hence, the significance of the H-maser
noise is expected to increase at higher frequencies and with better
weather conditions, while the CSO noise remains negligible in all
circumstances.

\begin{figure}
\includegraphics[width=8cm]{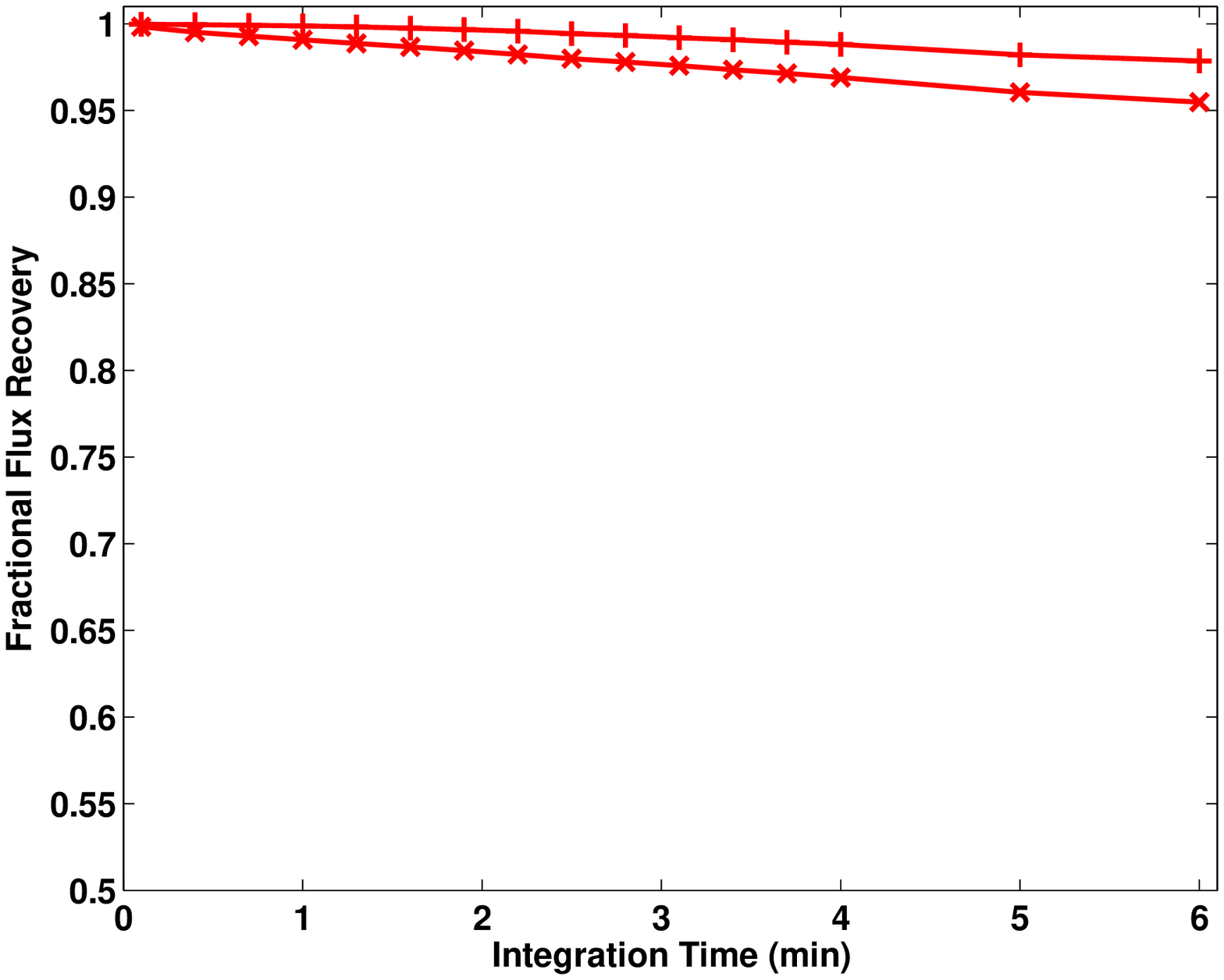}
\includegraphics[width=8cm]{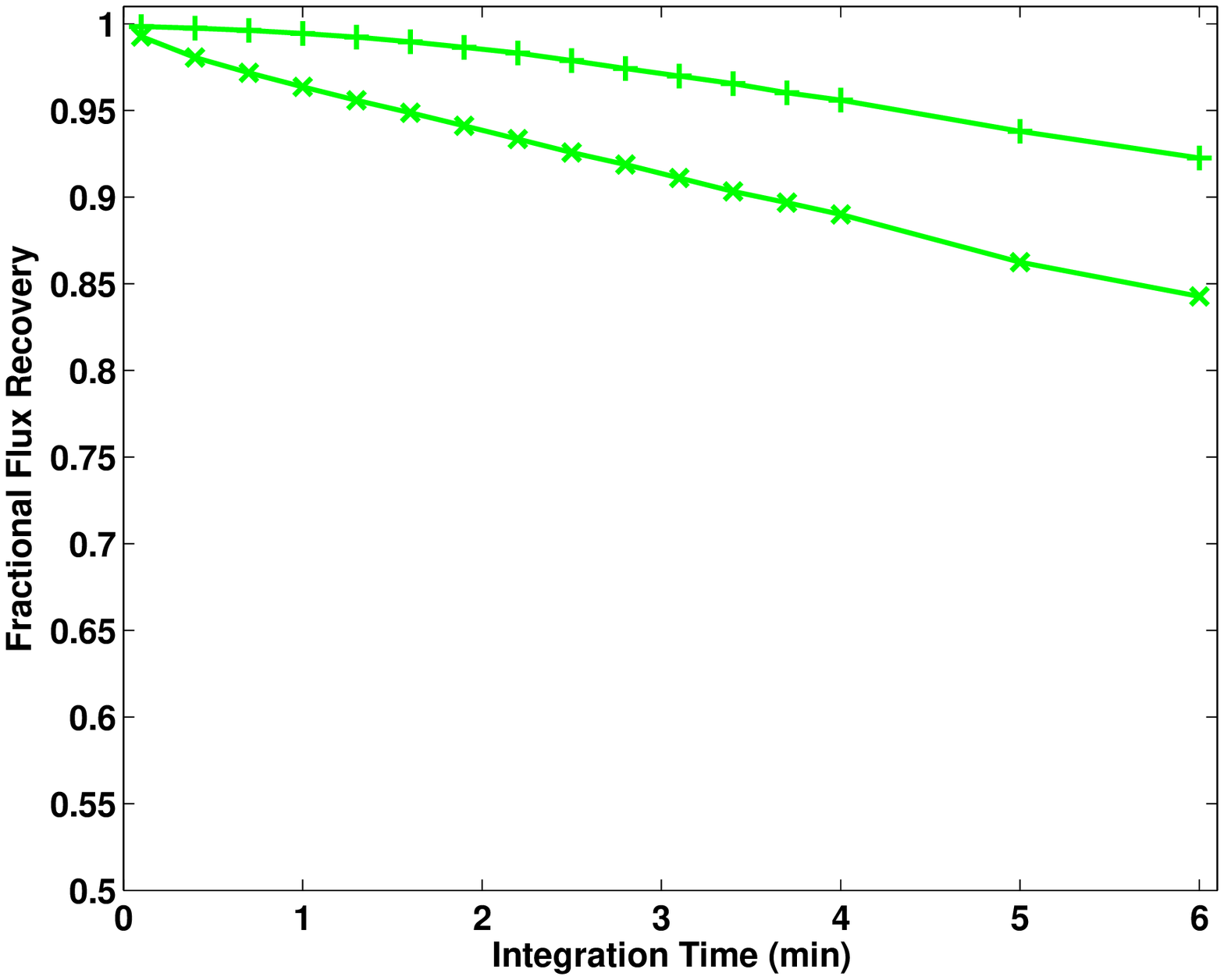}
\includegraphics[width=8cm]{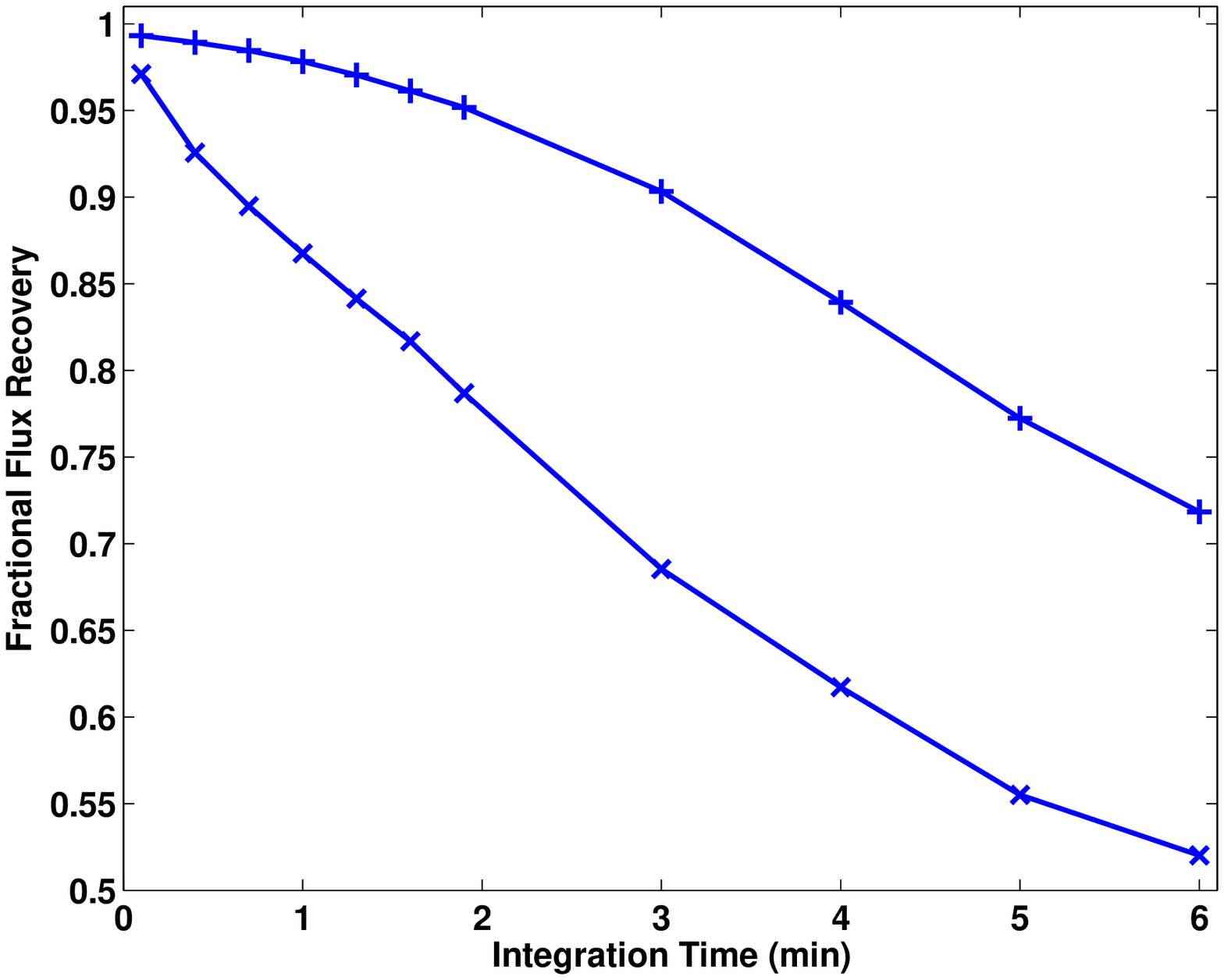}
\includegraphics[width=8cm]{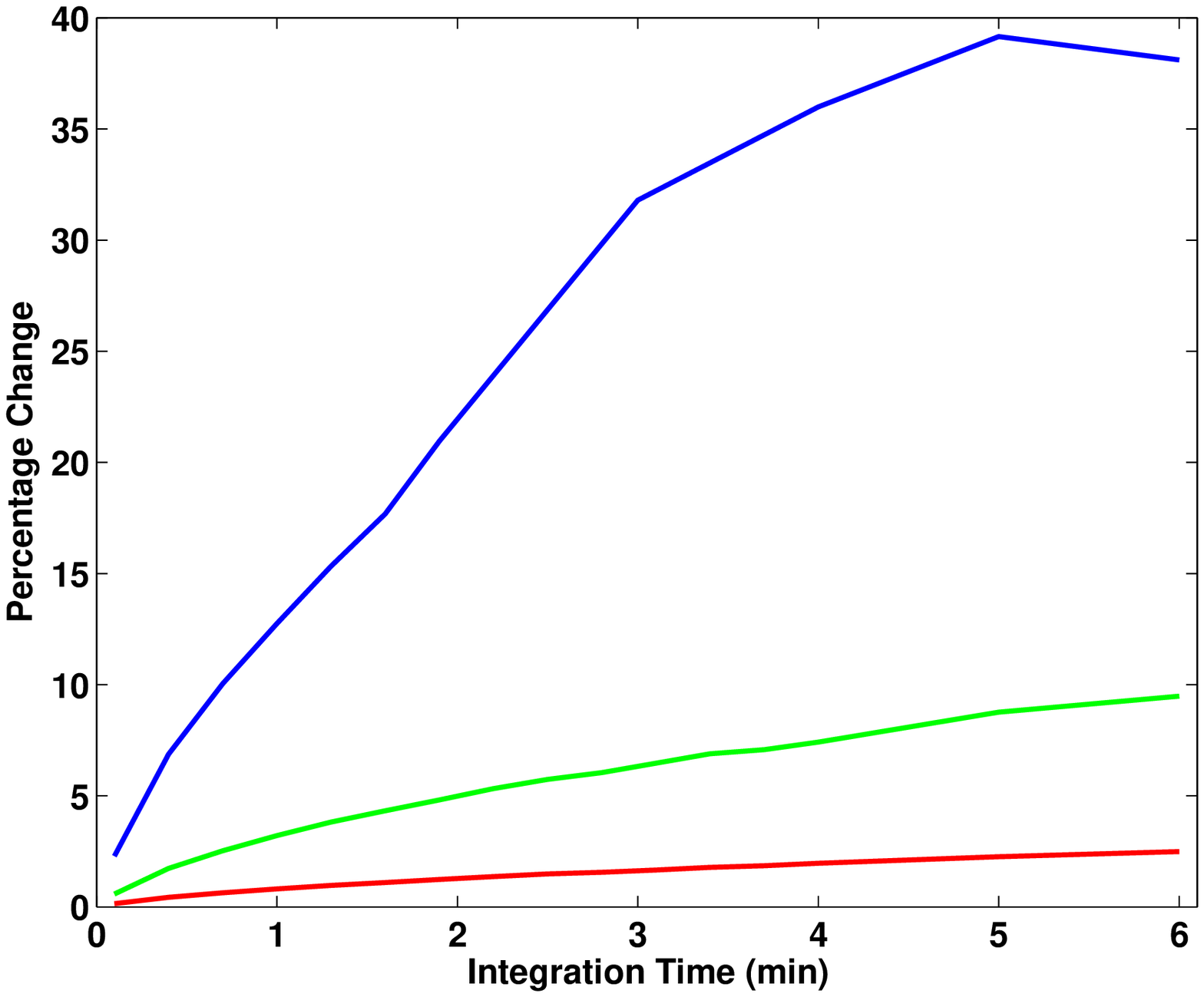}
\caption{The fractional flux recovered with WVR-corrected tropospheric atmospheric models plus clock (CSO `+'; H-maser `x') instabilities,
  versus integration time, at 86 ({\it red}, upper left), 175 ({\it green}, upper right) and 350 ({\it blue}, lower left) GHz.
  The {\sc rms} errors are smaller than the symbol size.
Additionally, the lower right plot shows the fractional increase in
coherence loss when using a H-maser in place of a CSO, versus
integration time, for the three frequencies. The frequencies run from the highest at the top to the
lowest at the bottom.}
\label{case3a}
\end{figure}

\begin{figure}
\includegraphics[width=8cm]{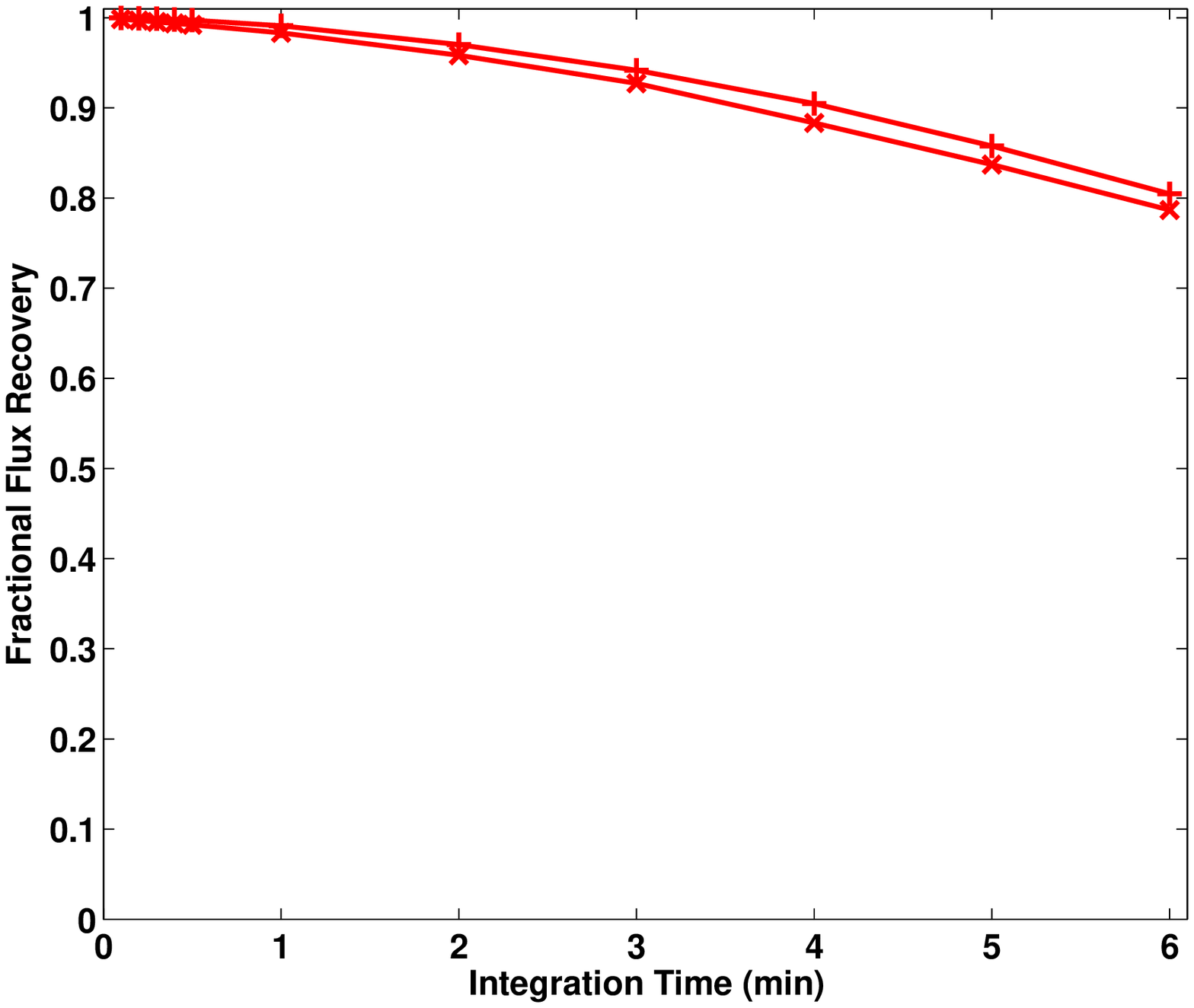}
\includegraphics[width=8cm]{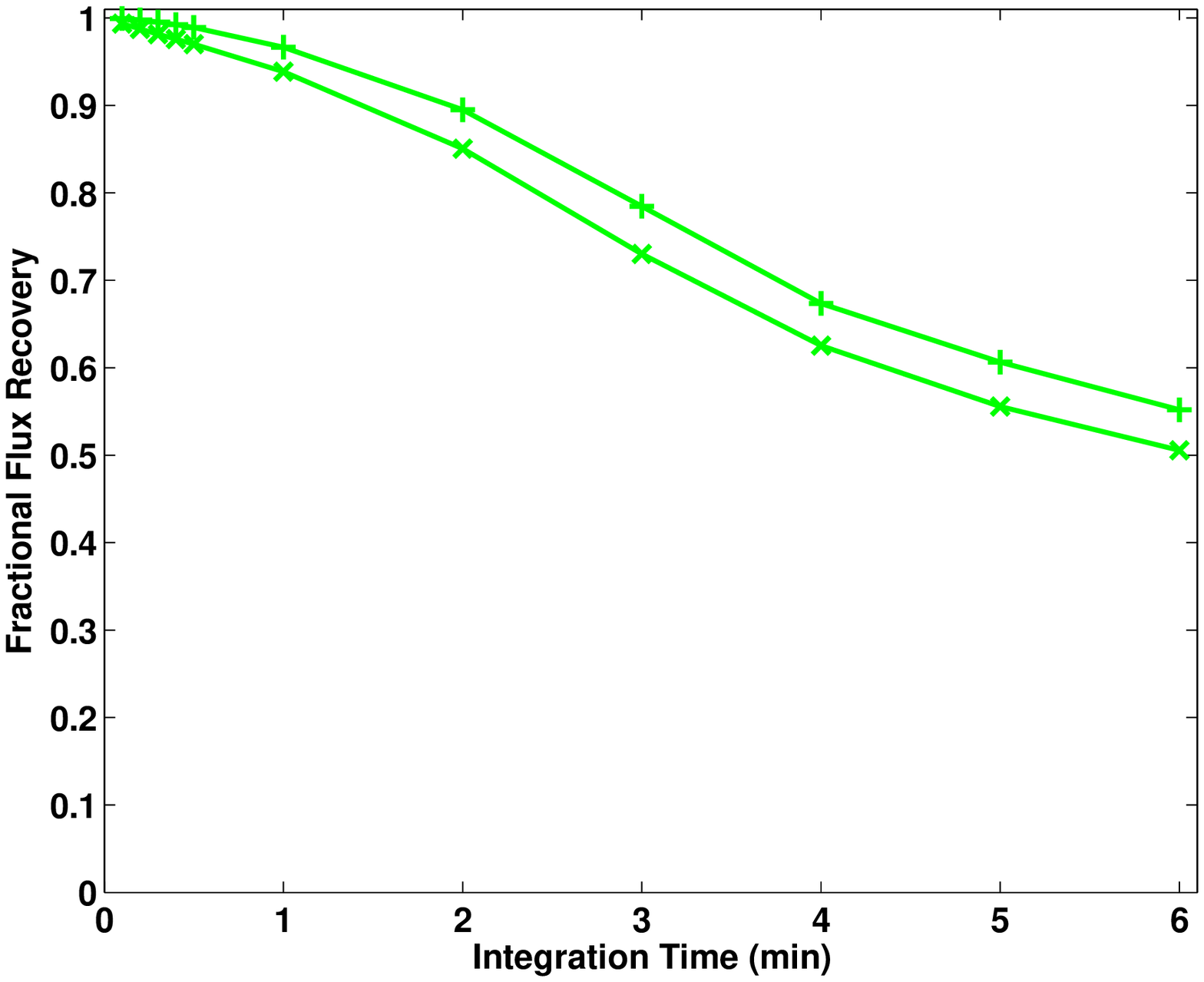}
\includegraphics[width=8cm]{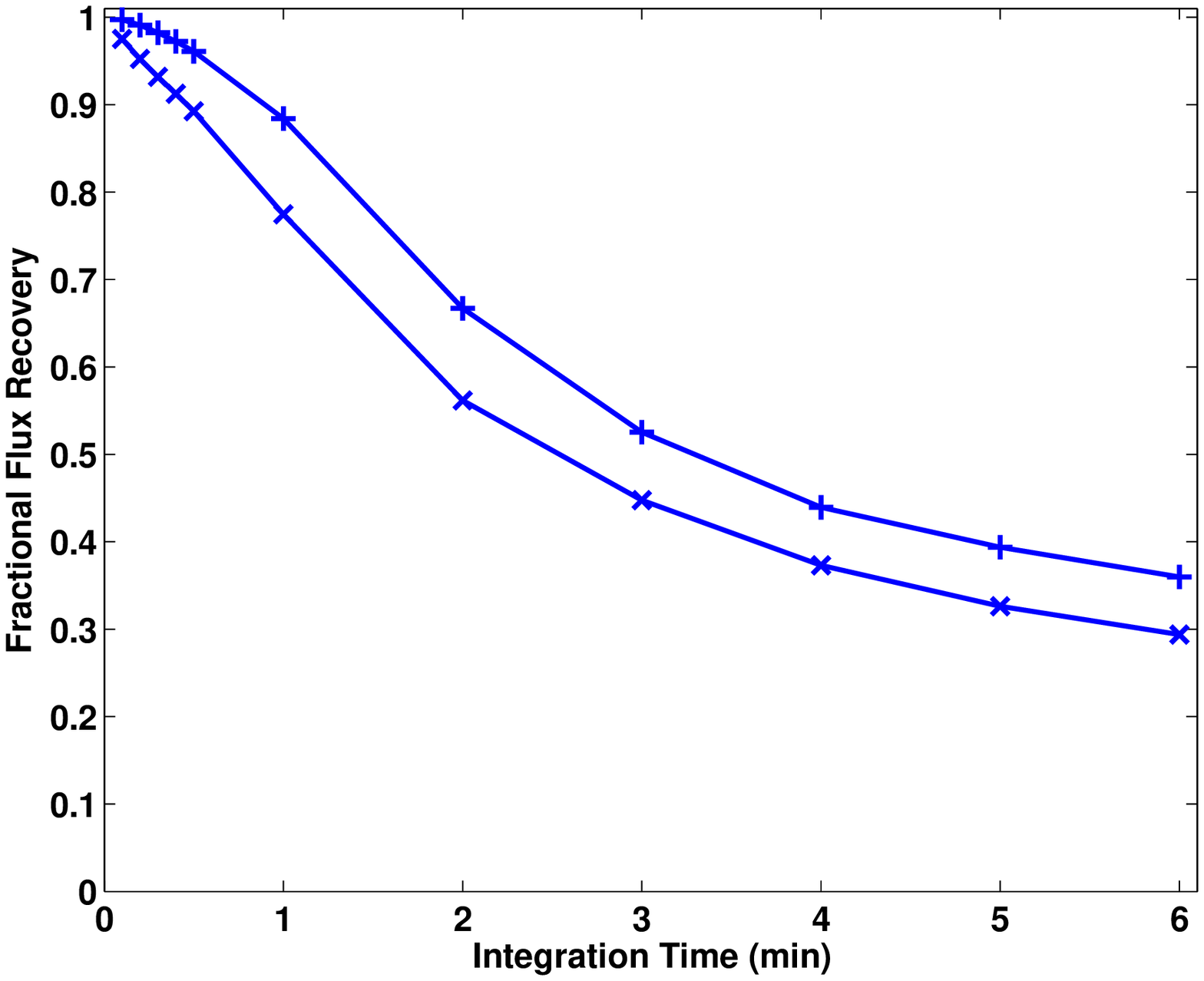}
\includegraphics[width=8cm]{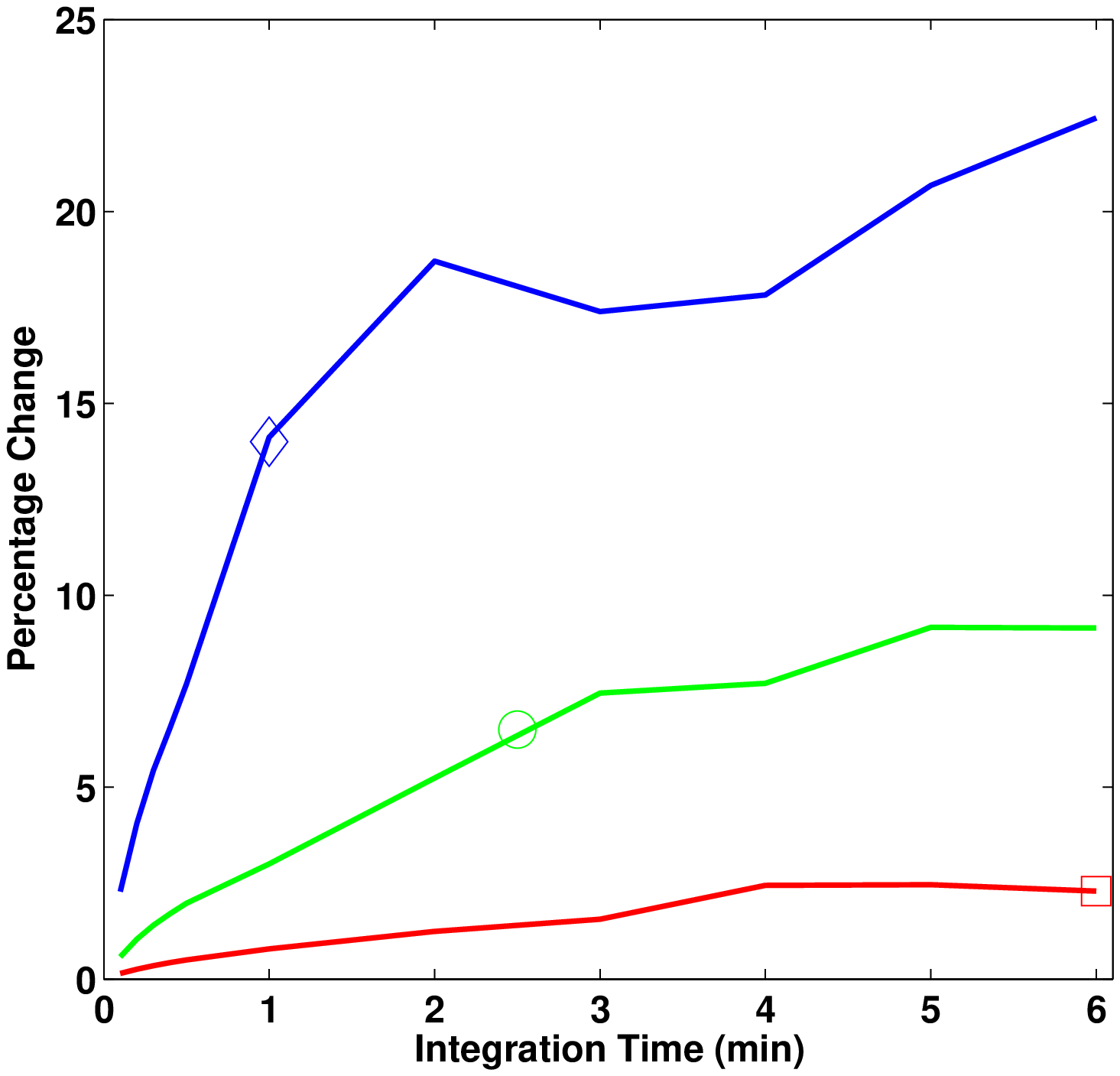}
\caption{The fractional flux recovered with Very Good (V) tropospheric  
  atmospheric models plus clock (CSO `+'; H-maser `x') instabilities,
  versus integration time, at 86 ({\it red}, upper left), 175 ({\it green}, upper right) and 350 ({\it blue}, lower left) GHz.
  The {\sc rms} errors are smaller than the symbol size.
  Additionally, the lower right plot shows the fractional increase in
coherence loss when using a H-maser in place of a CSO, versus
integration time, for the three frequencies as in Figure 5.
  For each frequency a symbol (a diamond for 350 GHz, a circle for 175
  GHz and a square for 86 GHz) marks the point corresponding to losses of
  20\% in the H-maser simulations.
%
}
\label{case3b}
\end{figure}

\begin{figure}
\includegraphics[width=8cm]{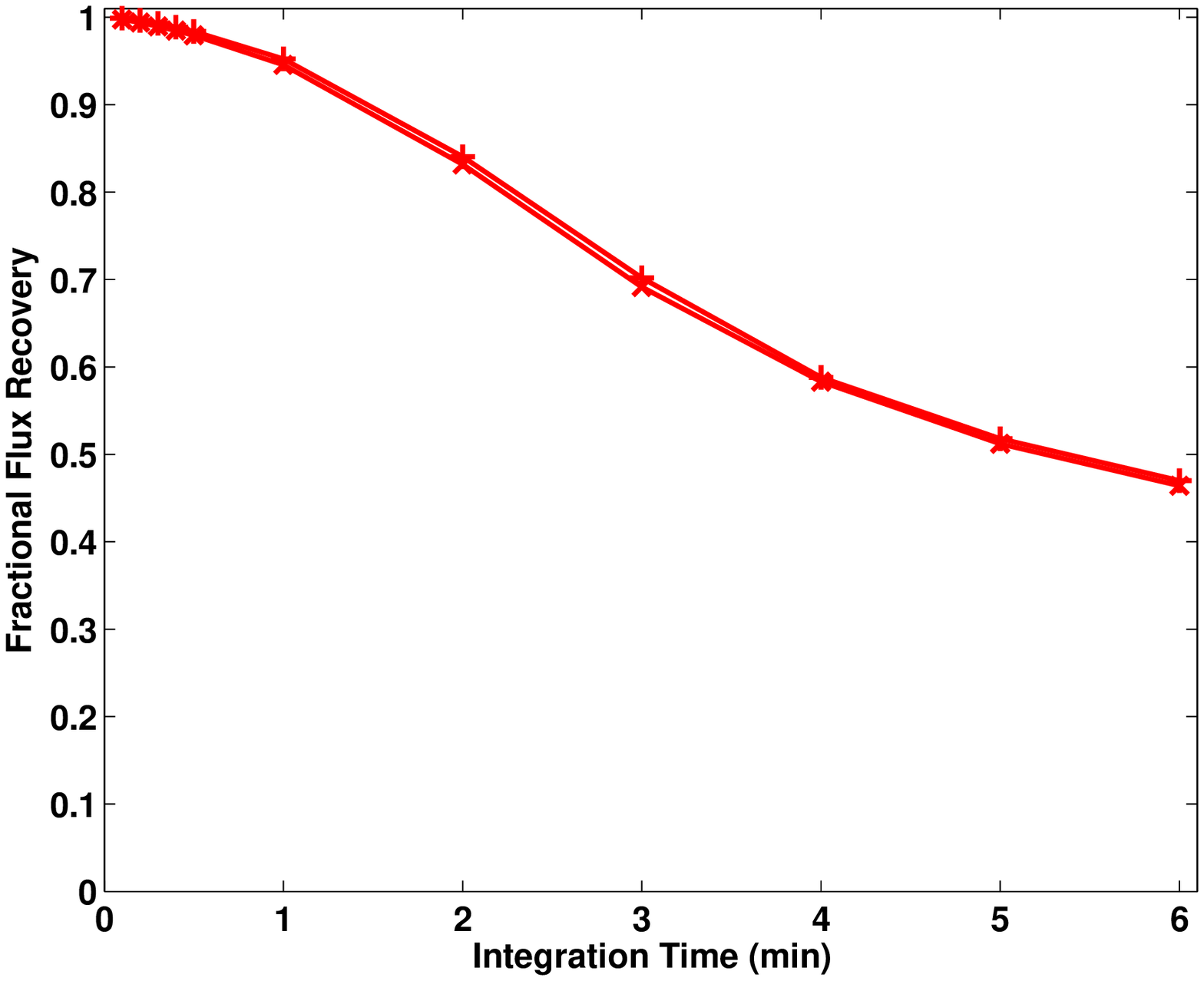}
\includegraphics[width=8cm]{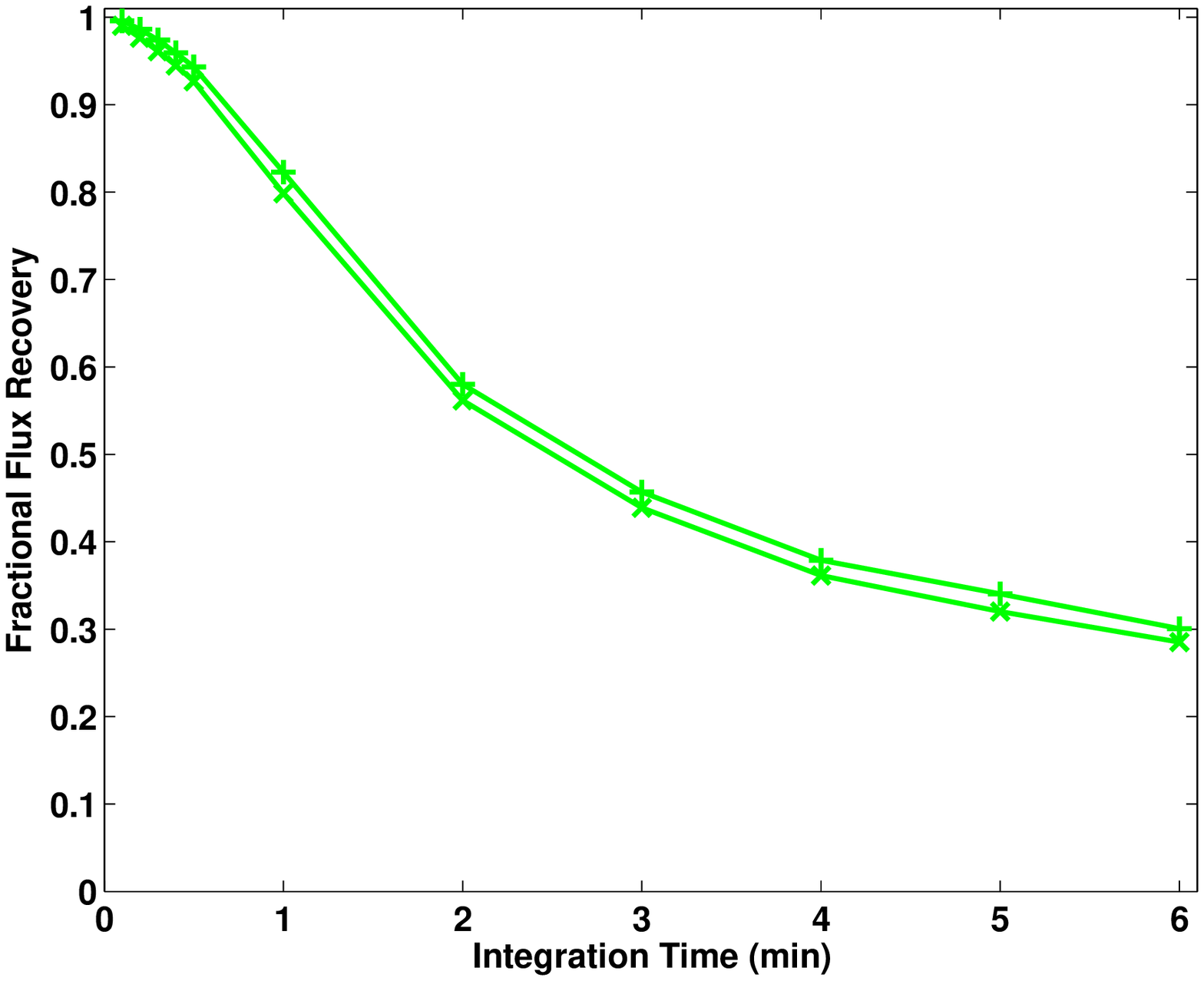}
\includegraphics[width=8cm]{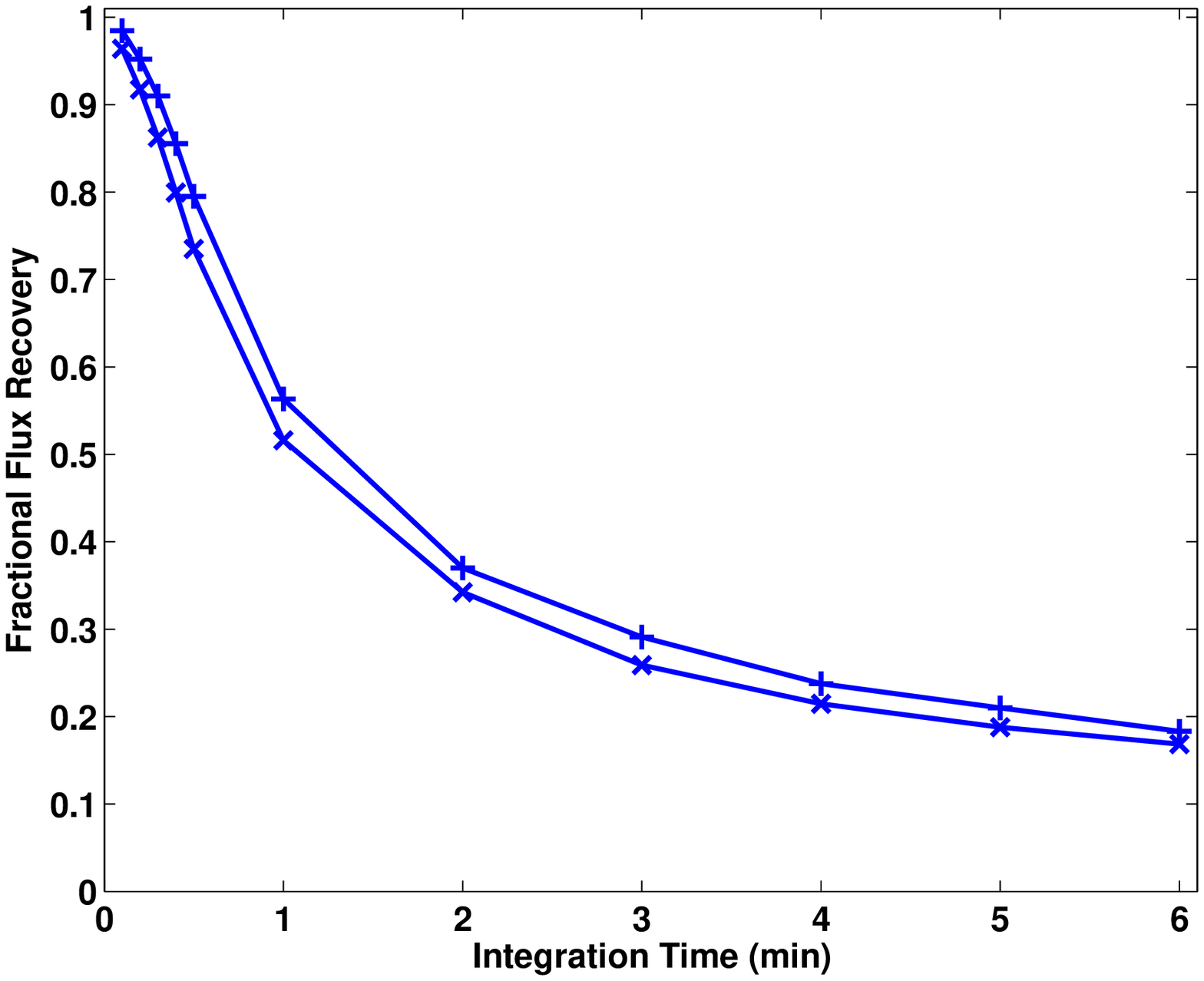}
\includegraphics[width=8cm]{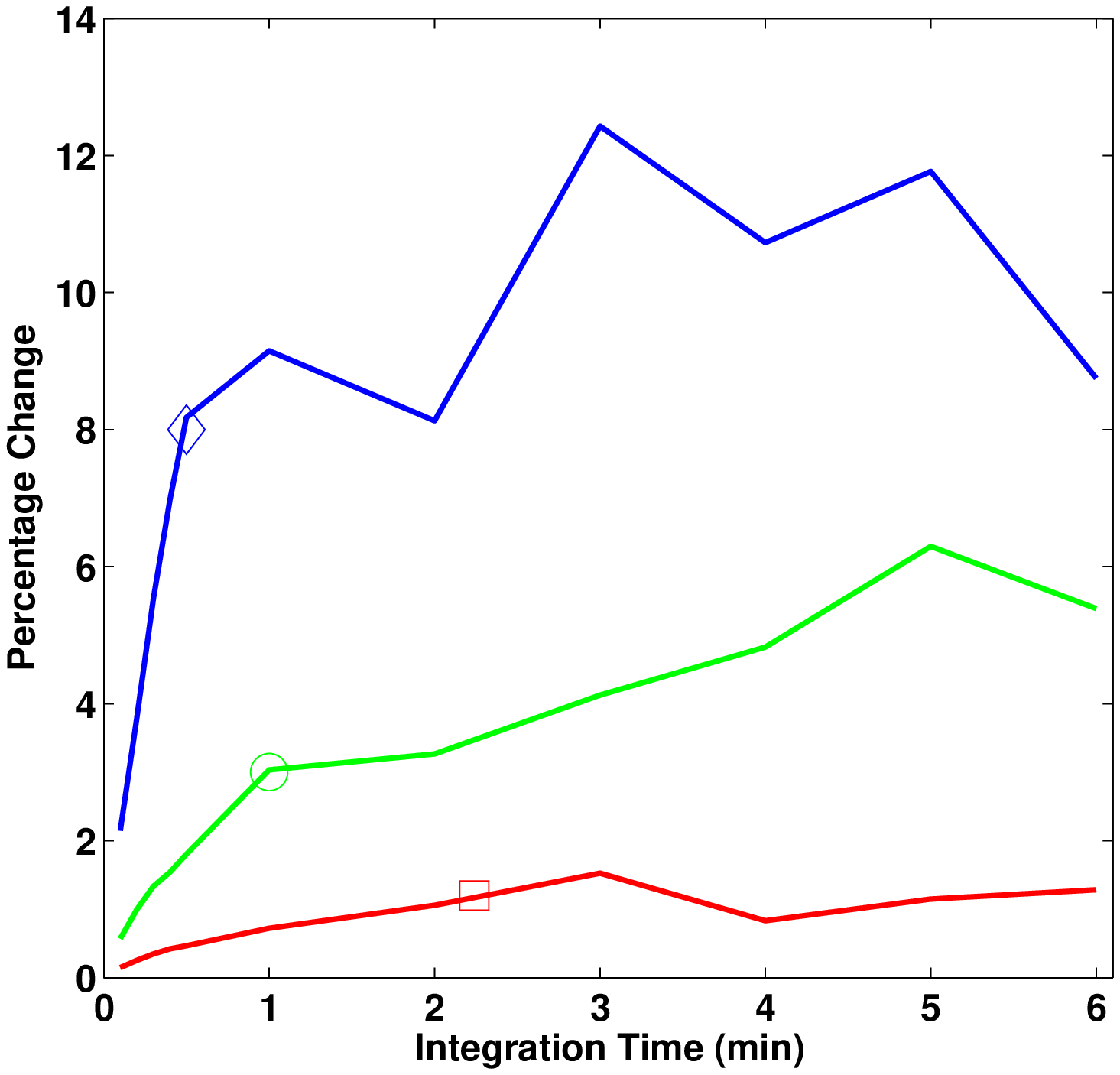}
\caption{The fractional flux recovered with Good (G) tropospheric  
  atmospheric models plus clock (CSO `+'; H-maser `x') instabilities,
  versus integration time, at 86 ({\it red}, upper left), 175 ({\it green}, upper right) and 350 ({\it blue}, lower left) GHz.
  The {\sc rms} errors are smaller than the symbol size.
  Additionally, the lower right plot shows the fractional increase in
coherence loss when using a H-maser in place of a CSO, versus
integration time, for the three frequencies as in Figure 5.
  For each frequency a symbol (a diamond for 350 GHz, a circle for 175
  GHz and a square for 86 GHz) marks the point corresponding to losses of
  20\% in the H-maser simulations.
%
}
\label{case3c}
\end{figure}

Figures \ref{case3a}, \ref{case3b} and \ref{case3c} show the 
results from realistic simulations with clock and atmospheric instabilities.
Figure \ref{case3a} shows superimposed the FFR
measured for the case of WVR-corrected atmospheres both with a H-maser and with a CSO 
at all frequencies.
Figures \ref{case3b} and \ref{case3c} are equivalent to Figure \ref{case3a}, but
for V and G weather conditions, respectively.
Additionally, Figures
\ref{case3a} to \ref{case3c} include the fractional change between the
FFR estimated from both simulation cases, with the H-maser and the CSO, at the
corresponding frequency and weather conditions, as a function of integration time.

\subsection{{\it Cases 4 and 5: Dual-Frequency Observations}}

This set of simulations explores the use of dual-frequency
observations to alleviate the limited coherence time problem in mm-VLBI. The results using simultaneous dual frequency
observations following FPT analysis are presented for the following
pairs of frequencies: 43/86GHz, 87/175 GHz, 175/350GHz.
Note that keeping an integer frequency
ratio reduces the problems related to phase ambiguities
in the analysis \citep{riojadodson11}.  

For each frequency pair, the coherence losses (at the higher frequency) 
are greatly reduced by a two step procedure. First, using
the scaled `first step' calibration from the lower frequency 
to provide compensation for the non-dispersive fast
fluctuations, combined with `second step' self-calibration analysis with a much
longer scan span, which removes the remaining dispersive residuals. The non-dispersive
terms comprise the tropospheric and frequency standard instabilities;
among the dispersive, the ionospheric and instrumental terms.
The residual ionospheric contributions in the higher frequency
FPT-calibrated dataset are those at the lower frequency, magnified by
the frequency ratio.  Even for the highest frequency pairs these are
many cycles of phase that would prevent imaging. The
coherence timescales have been observed to be $\sim$30 minutes at
86\,GHz and would be expected to be even longer at higher frequencies.

\begin{figure}
\includegraphics[width=7cm]{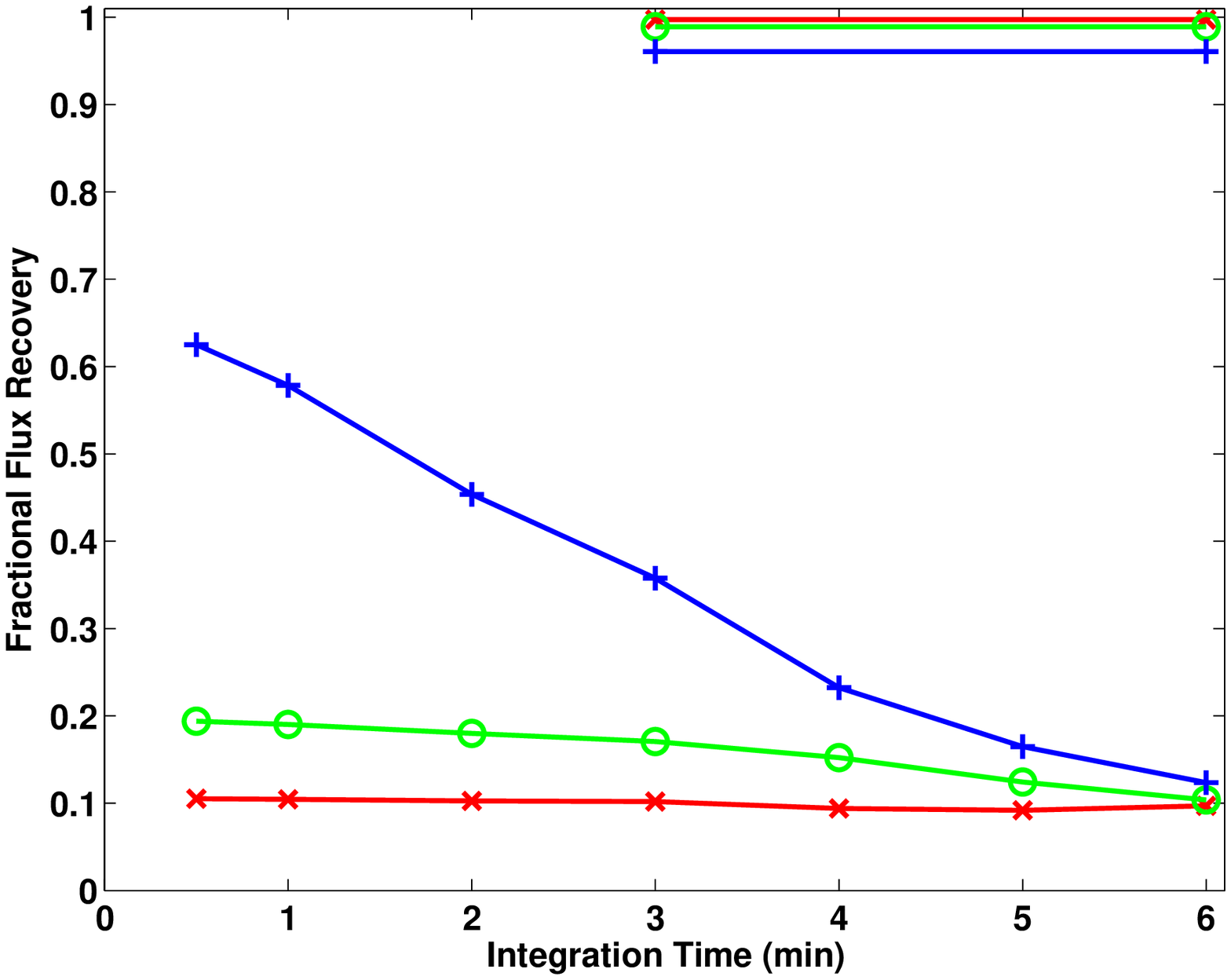}
\includegraphics[width=7cm]{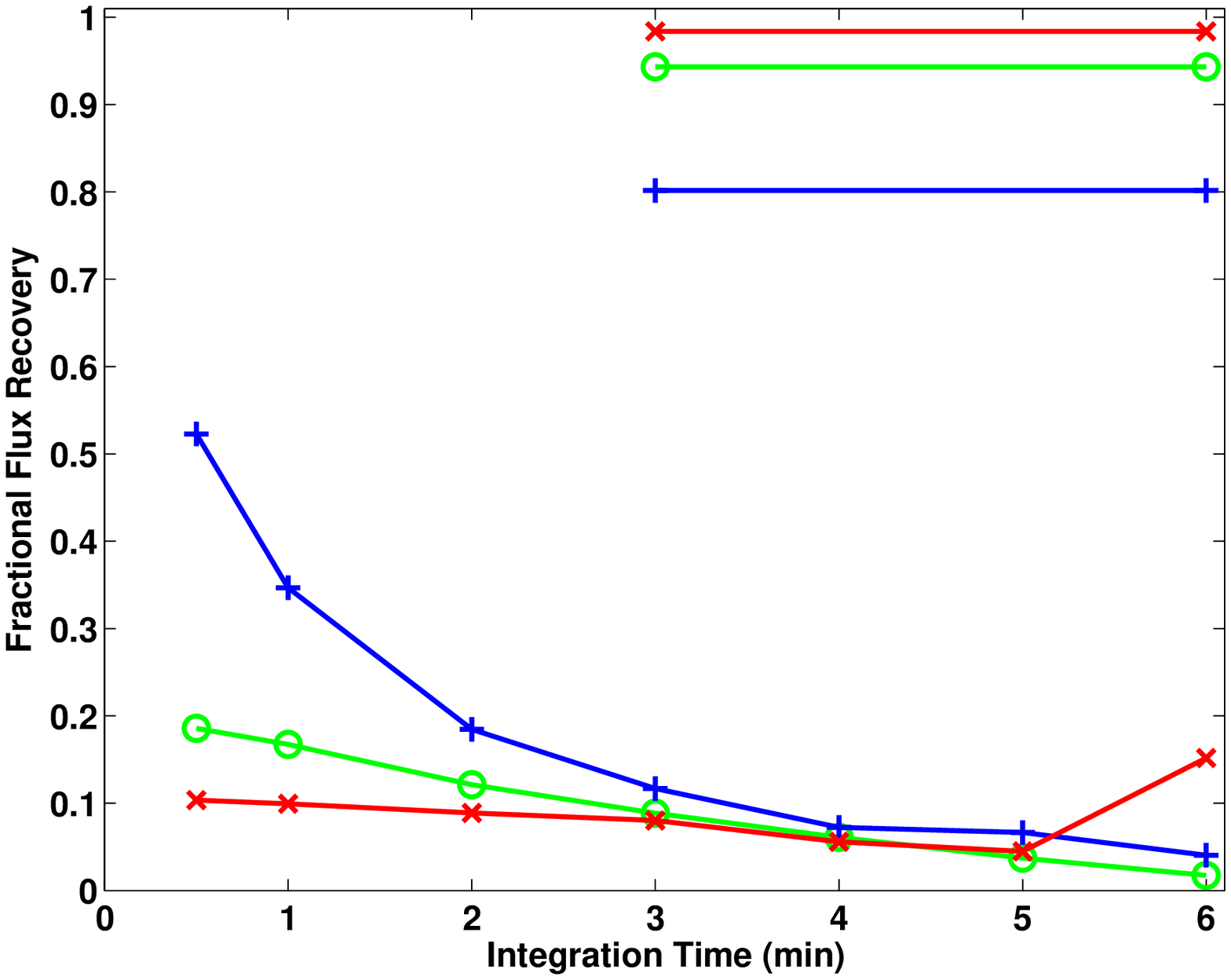}
\caption{The fractional flux recovered versus integration time for the Frequency Phase Transfer (FPT) analysis 
of realistic simultaneous observations of pairs of frequencies, at
43/86 GHz ({\it red}; cross), 87/175 GHz 
({\it green}; circle), and 175/350 GHz ({\it blue}; plus) under Very Good ({\it Left}), and Good ({\it Right}) weather 
conditions.
The {\sc rms} errors are smaller than the symbol size.
Also shown in the top right are the values obtained using FPT plus a
further self calibration iteration, following the 
same color and symbol codes.}
\label{case4}
\end{figure}

Figure \ref{case4} shows the FFR values measured from the images
at the higher frequencies, using the scaled calibration estimated
from the lower (i.e. FPT), for each pair of frequencies,
and for a range of scan integration times ({\it case 4}) for V and G weathers. Note that
the FFR values are larger for the higher frequency pairs, 
due to the smaller ionospheric residual errors in the FPT analysis 
at higher frequencies.
The results from a hybrid analysis comprising a FPT pre-calibration
(with 30 sec integration time) followed by a run of self-calibration
at the high frequency (with 3 and 6 minutes integration time),
for each pair of frequencies
({\it  case 5}) are also shown in the same Figure.

\section{Discussion}

\subsection{Comparative Performances}

VLBI traditionally uses H-masers as frequency
standards. Nevertheless, the short coherence times in observations at
mm-VLBI can turn the H-maser instabilities into
a significant limitation.  We have carried out simulation
studies to explore and compare the performance of H-masers and ultra
stable CSOs in mm-VLBI observations, 
and quantify the impact with respect to that from the tropospheric 
fluctuations, under a variety of circumstances. 
Our simulations show that, in most situations, 
the tropospheric fluctuations constitute the
dominant contribution to the 
coherence losses at mm and sub-mm wavelength observations; 
its 
impact increases with the observing frequency and/or worsening weather conditions,
which in turn limits the coherence time and sensitivity.  
Hence, selecting sites with stable weather conditions is an essential step towards increasing
the sensitivity of high frequency observations, as is well known.
The impact of the coherence losses resulting from instabilities in the
time standards in high frequency VLBI has been less explored. Its
impact can be expected to be larger at higher frequencies.  
Our simulations show that at the highest
frequencies, the losses induced by the H-maser instabilities are comparable to those from quality
tropospheric conditions, for timescales comparable to the coherence times.
For comparison, the corresponding losses for CSO instabilities are negligible ($<$0.5\%).
Hence, benefits from replacing the H-maser with a CSO for observations
at the highest frequencies can be expected.  In the other 
cases, the tropospheric fluctuations are the limiting
contribution and there is little difference whether ones uses a H-maser or a CSO. \\
Figures \ref{case3b} and \ref{case3c}
show the percent changes between the coherence losses for the H-maser and
the CSO simulations as a function of integration time, in V and G
weather conditions.
For example, those are $\sim 2$, 7 and 14\% at 86, 175 and 350 GHz,
respectively, for $\sim 20\%$ coherence loss in the
H-maser simulations with V weather conditions; and $\sim 1$, 3,
8\% with G weather conditions.
We conclude that the benefits, estimated by the reduction in coherence losses for the same
integration time, 
of replacing the H-maser with a CSO are significant and 
$\ge 10\%$ for frequencies above 175 GHz (without WVR).

\subsection{CSOs in VLBI Sites with Collocated WVRs}

The use of co-located WVR in VLBI sites has the potential to greatly
compensate the phase fluctuations induced by the tropospheric
propagation into the astronomical signals.  The measured effectiveness
of the ALMA 183 GHz WVRs promises to provide a breakthrough for
mm-VLBI observations \citep{honma_alma,matsushita_spie} by enabling
longer coherence times and hence higher sensitivity even at the
highest frequencies and under a wider range of weather conditions.
Its effect is roughly equivalent to an upgrade in the array weather
conditions, which is dominated by the site with the poorest
conditions, and will lead to the realization of a sub-mm VLBI array
comprising multiple telescopes with ALMA-type weather conditions
(i.e. upgraded from G to V).

 The superior tropospheric compensation provided by
co-located WVRs increases the significance of the H-maser
instabilities and turn them into the dominant limitation in a much
wider range of circumstances than shown in our simulations without
WVR: i.e. over longer timescales, under more normal weather
conditions, and for observations at lower frequencies.  
Maximum benefits are to be obtained from the combined
operation of WVR and CSO at sub-mm VLBI sites.

\begin{figure}
\includegraphics[width=10cm]{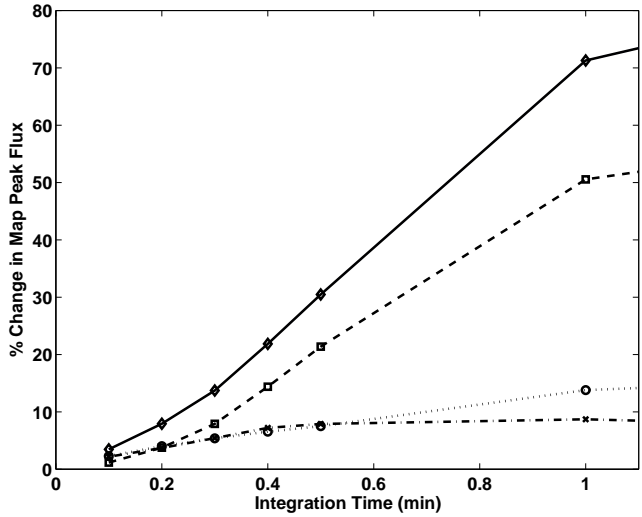}
\caption{ 
Comparative performances (i.e. percent change of coherence losses) of VLBI observations at 350 GHz versus 
integration times under different combinations of clocks and weather conditions: 
1) impact of clock, using CSO instead of H-maser: 
CSO/G vs H-maser/G (dot-dash line) and CSO/V vs H-maser/V (dotted line); 
2) impact of weather, using WVR to upgrade G to V conditions: H-maser/V vs H-maser/G (dash line); 
3) combined effect of ultra stable clock and WVR: CSO/V vs. H-maser/G (solid line).}
\label{350upgrade}
\end{figure}

Figure \ref{350upgrade} 
shows the relative impact estimated for improved weather
conditions as a result of using WVRs, 
for the more precise frequency standard,
and for the combined effect 
of WVR and CSO 
in observations at 350 GHz.  
We estimate $\sim 10-20\%$ improved performance (i.e. reduced coherence loss) 
for observations under G weather conditions in a range of integration times
from ten to a few tens of seconds,
the typical coherence times found in VLBI observations at 1.3 mm 
with H-masers and no WVR corrections \citep{Doeleman11}.
Note that longer coherence times are expected as a result of the WVR atmospheric 
phase compensation and the ultra stable CSO, which result in an increase of the 
sensitivity.
For comparison, at 175 GHz, a similar improvement is estimated for 0.5 to 1 minute timescales.
Hence, we estimate significant benefits in combining WVRs and CSOs for VLBI 
observations at frequencies $\ge 175$ GHz.  \\

\subsection {Increased Sensitivity Derived from Ultra Stable Frequency Standards.}

\begin{figure}
\includegraphics[width=10cm]{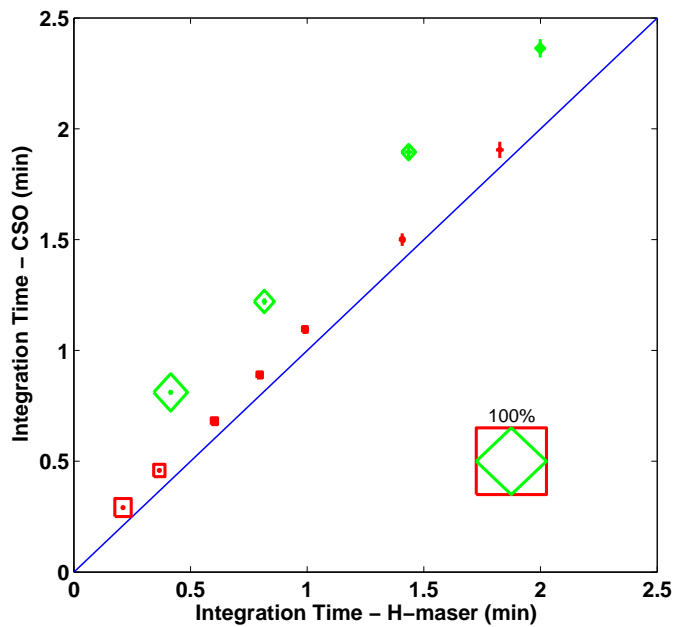}

\includegraphics[width=10cm]{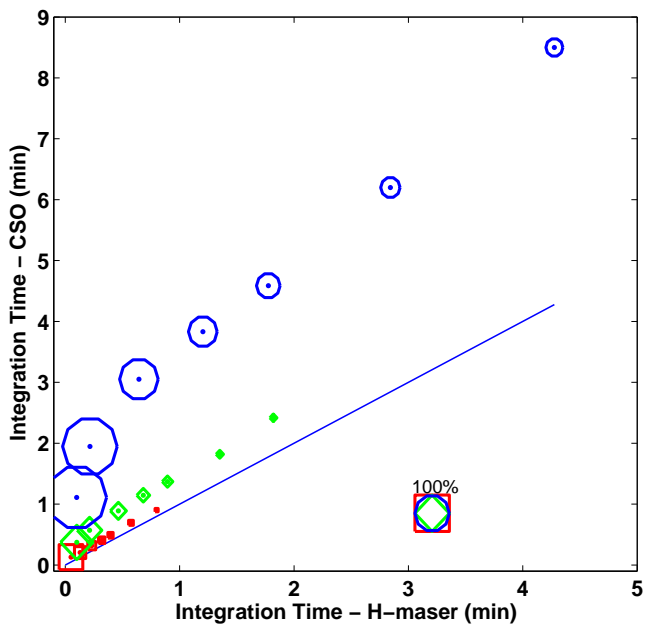}
\caption{Comparison of integration times for matching coherence losses
  in complete simulations with CSO ({\it y-axis}) and H-maser ({\it x-axis}), under
  Very good (green diamond) and Good (red square)) weather conditions at 175 GHz ({\it Upper}), 
  and additionally for WVR-corrected atmospheres (blue circle) at 
  350 GHz ({\it Lower}). 
   Range of coherence losses displayed is: 2.5, 5, 10, 15, 20, 30, 40\%; the points line up in 
   the same order, with increasing distance from the origin. For each point, the size of the symbol
  corresponds to the log of the increased sensitivity value estimated from 
  the use of the CSO compared to the H-maser for the corresponding coherence loss.}
\label{tint-tint}
\end{figure}

In this section we take a different approach to compare the performances of H-maser
and CSO in VLBI observations, in terms of sensitivity. Rather than comparing the 
coherence losses as a function of the integration time, here we compare the 
integration times that lead to identical losses, with the CSO and H-maser
respectively, as a function of the coherence loss.
The increase in sensitivity can be calculated as the square root of their ratio.
Figure \ref{tint-tint} shows the necessary 
CSO versus H-maser integration times for selected coherence losses between
2.5--40\%, measured from simulations at 175 and 350 GHz 
under a variety of weather conditions.
The locus of points for each
of the frequency-weather combinations share similarities:
the points are distributed, approximately, along lines parallel to y=x
(the bottom-left to top-right diagonal); the total span 
is shorter for higher frequencies and worse weather conditions;
the offset of each dataset from the diagonal increases with the 
quality of the weather conditions.
For each point this offset is the difference between the integration
times, for the CSO and H-maser cases, to reach an equal coherence loss. 
The sensitivity gain is the square root of the ratio of these integration times, as
is indicated in the plot by the symbol size.
Following this criterion, the increased sensitivity obtained from the CSO with
respect to the H-maser in observations at 175 GHz with best (V) weather conditions 
are estimated to be 9\% and 13\%, for coherence losses of 20\% and 10\%, 
respectively; the equivalent values for good (G) weather conditions are 5\% and 8\%,
respectively. 
At 350 GHz, the estimated increased sensitivity values are 22\% and 41\%, respectively,
in V weather conditions. For G weather these are 11\% and
18\%.  Maximum values of increased sensitivities are estimated for the case of WVR-corrected 
atmospheres (W) at 350 GHz, equal to 60\% and 120\%, for coherence losses of 20\% and 10\%, 
respectively. In all cases, the estimated values for increased sensitivities 
are larger for lower acceptable coherence losses. 
We conclude that there are significant sensitivity benefits, resulting from increased integration time,  
from replacing the H-maser with a ultra stable CSO, for observations at frequencies $\ge 175$ GHz, 
and that maximum benefits are obtained from using 
collocated WVRs to compensate for tropospheric fluctuations. \\

\subsection {Application of CSOs in Space VLBI}

Our simulations show the tendency for increasingly significant
benefits to be gained from using CSOs (instead of H-masers) in VLBI
observations at higher frequencies and under best weather conditions.
Following these lines, space mm-VLBI, with no atmospheric propagation
issues (at least for the space segment), provides the optimum case.
\citet{asaki_leo09} has proposed a low earth orbit satellite
constellation, which provides excellent uv-coverage and highest
angular resolution. This would observe at 350 GHz, focusing on studies
of black hole environments.
Based on our studies those advantages can be expected to be further increased, in terms of sensitivity, when
combined with CSOs as time standards.

\subsection {Dual-Frequency Observations and Analysis Strategies} 

Our simulations comprise the exploration of dual-frequency observations and 
phase transfer analysis to be used in mm-VLBI.
Simultaneous dual-frequency observations are most desirable in the high frequency regime, in order to prevent 
the losses associated with switching.
Our estimates show that even at 350 GHz, very small coherence 
losses (5\% for V) can be obtained with integration times of many minutes, if
the data are pre-calibrated using observations at 175 GHz.  
An alternative option to the self-calibration step is to include
dual-frequency observations of a second source, 
with interleaving observations every few minutes;
this is the basis of the Source Frequency Phase Referencing analysis \citep{riojadodson11}, which provides spectral
astrometry between the two frequency bands along with the increased
coherence time for detection of weak sources.  In both scenarios, the
impact of tropospheric and clock instabilities, being of 
non-dispersive nature, are mitigated by the
pre-calibration derived from the lower frequency.
But fluctuations on timescales shorter than the pre-calibration integration
times will remain. 
Here too benefits are to be gained by installation of WVR, to 
continuously correct for those errors arising from the troposphere. 
The benefits in terms of sensitivity are very significant, but should be considered along with 
the downside of the magnification of the noise, which sets limits on
the minimum signal-to-noise ratio for
observations of weak sources.

\subsection {Extrapolation of Results to Higher Frequencies} 

We can extrapolate the results from our simulations at 86, 175 and 350
GHz to even higher frequencies using the trends shown in Figure
\ref{tint-tint}. At frequencies $> 350$ GHz, with best (V) weather
conditions, the locus of points corresponding to the same range of
coherence losses will spread along a line whose length will be shorter
than that shown in the plot at 350 GHz, and the corresponding 
percentage increase in the sensitivities by use of CSOs with respect to H-masers will be
larger. Using co-located WVRs, which will probably be mandatory at
this frequency regime, will result in a vertical shift of this line in
the plot, and in a net increase in the benefits of replacing 
the H-maser with a CSO. 
Hence, these results make a compelling case for use of CSOs 
in VLBI observations at the highest frequencies, beyond those in our simulations.

\subsection{Possible Improvement of CSO Time Standards} 

The current CSO performance has potential for improvement.
Both the 100 MHz and 10 MHz signals are synthesised from
the microwave frequency of the oscillator which is determined by the natural resonance of the sapphire crystal (11.2-GHz).  Due to the
intrinsic phase noise of the hardware components that synthesize these
frequencies, the 100 MHz is produced with much less degradation of
the signal quality. Thus it is the recommended reference signal and is the frequency that has been used in
these studies. Ongoing research also indicates that 1 GHz signals can be
synthesized with even less degradation and such reference signals can
now be transmitted over long distances with fiber techniques
(e.g. \citet{wilcox-09}). Thus the time standards no longer need to be on-site. 
Hence, should it be necessary in the future
to further improve the CSO reference for the very highest observation
frequencies there is still room for improvement. However we note that
even in the best conditions, with WVR corrections, the atmospheric
contributions are much greater than those from the CSO. 

\section{Conclusions}

We have simulated realistic mm-VLBI data to address the
importance of improving the frequency standards. We have extended the
simulation tool ARIS to include H-maser and CSO instabilities,
and allow for best quality and WVR-corrected atmospheric conditions.
These simulated datasets reproduce well the expected behaviours from
observational data.
We have also explored the potential of dual-frequency observations to increase
the coherence time at the higher frequencies in mm-VLBI observations.
We conclude that, even at the highest frequencies, further calibration, 
such as Self-Calibration or SFPR, is required to produce useful results.

We have compared the coherence losses as a function of integration time
using the simulations with H-maser and CSO, also we have estimated the improved
sensitivity resulting from the increased integration time for a given allowable signal loss.
We find that CSOs make a significant improvement at frequencies $\ge$175 GHz, 
under the best weather conditions. 
Using colocated WVRs to compensate for tropospheric
fluctuations, we find the improvement to be even greater, and for a wider range of weather conditions.


\end{document}